\documentclass[noshowpacs,amsmath,twocolumn,superscriptaddress,10pt,aps,prb]{revtex4-1} 
\usepackage{setspace}
\usepackage{amsmath}
\usepackage{breqn}
\usepackage{graphicx}
\usepackage[nearskip,margin = 0pt,caption=false]{subfig}

\usepackage{verbatim}
\usepackage{amsfonts}
\usepackage{amssymb}
\usepackage{epstopdf}
\usepackage{lipsum}
\usepackage{siunitx}
\usepackage{xcolor}
\usepackage{array}
\usepackage[resetlabels,labeled]{multibib}
\usepackage[normalem]{ulem}
\DeclareGraphicsExtensions{.pdf,.eps,.png,.jpg,.mps}
\usepackage{etoolbox}

\begin{document}

\title{Hybrid Kerr-electro-optic frequency combs on thin-film lithium niobate}
\author{Yunxiang Song$^{1,*}$, Yaowen Hu$^1$, Marko Lon\v{c}ar$^{1,*}$, Kiyoul Yang$^{1,*}$\\
\vspace{+0.05 in}
$^1$John A. Paulson School of Engineering and Applied Sciences, Harvard University, Cambridge, MA, USA.\\}

\begin{abstract}
\noindent 
\color{black} 
\textbf{
Optical frequency combs are indispensable links between the optical and microwave domains, enabling a wide range of applications including precision spectroscopy\cite{udem2002optical,diddams2020optical,suh2016microresonator,coddington2016dual,picque2019frequency}, ultrastable frequency generation\cite{spencer2018optical, fortier2011generation,li2014electro,xie2017photonic,liu2020photonic,tetsumoto2021optically}, and timekeeping\cite{papp2014microresonator,newman2019architecture}. Chip-scale integration miniaturizes bulk implementations onto photonic chips, offering highly compact, stable, and power-efficient frequency comb sources. 
State of the art integrated frequency comb sources are based on resonantly-enhanced Kerr effect\cite{herr2014temporal,kippenberg2018dissipative} and, more recently, on electro-optic effect\cite{zhang2019broadband,rueda2019resonant,hu2022high,yu2021femtosecond,boes2023lithium}. While the former can routinely reach octave-spanning bandwidths and the latter feature microwave-rate spacings, achieving both in the same material platform has been challenging. Here, we leverage both strong Kerr nonlinearity and efficient electro-optic phase modulation available in the ultralow-loss thin-film lithium niobate photonic platform, to demonstrate a hybrid Kerr-electro-optic frequency comb with stabilized spacing. In our approach, a dissipative Kerr soliton is first generated, and then electro-optic division is used to realize a frequency comb with 2,589 comb lines spaced by 29.308 GHz and spanning 75.9 THz (588 nm) end-to-end. Further, we demonstrate electronic stabilization and control of the soliton spacing, naturally facilitated by our approach. The broadband, microwave-rate comb in this work overcomes the spacing-span tradeoff that exists in all integrated frequency comb sources, and paves the way towards chip-scale solutions for complex tasks such as laser spectroscopy covering multiple bands\cite{shi2023frequency}, micro- and millimeter-wave generation\cite{fortier2011generation,koenig2013wireless,li2014electro,liu2020photonic,tetsumoto2021optically,wang2021towards}, and massively parallel optical communications\cite{marin2017microresonator,Oxenlowe:2022:NaturePhotonics,Vuckovic:2022:NatureCommun,rizzo2023massively}.}
\end{abstract}

\maketitle
 
\begin{figure*}[t!]
\centering
\includegraphics[width=\linewidth]{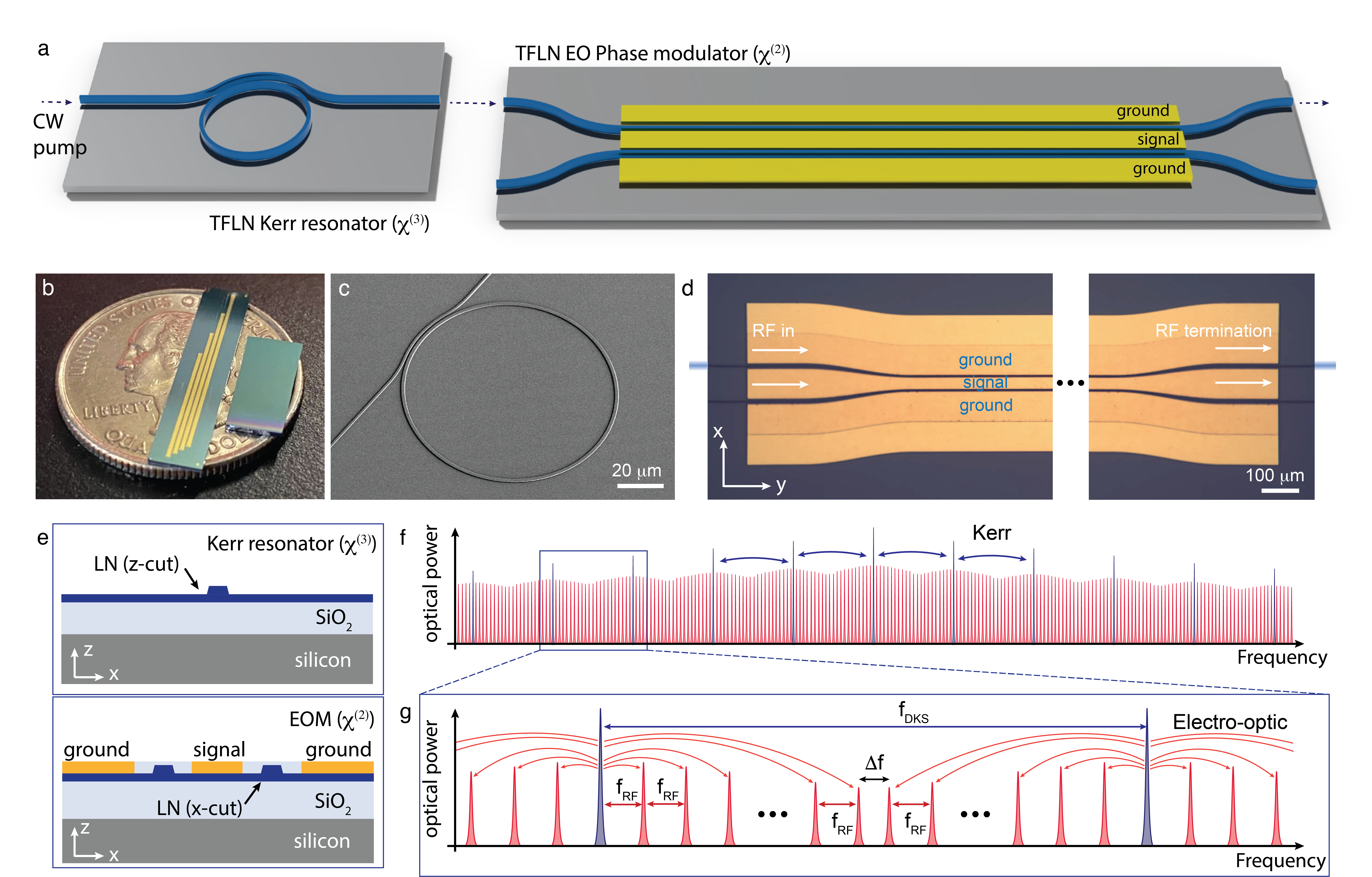}
\caption{\label{fig:Fig1}\textbf{Concept of hybrid Kerr-electro-optic frequency comb.} \textbf{a} 3-D illustration of the hybrid Kerr-electro-optic frequency comb system, consisting of a dissipative Kerr soliton microresonator chip and an electro-optic phase modulator chip. A continuous-wave (CW) optical frequency initiates a THz-rate soliton frequency comb which then undergoes electro-optic division into a microwave-rate hybrid comb after passing through the modulator. \textbf{b} Photograph showing TFLN photonic chips with electro-optic phase modulator array (left) and Kerr resonators (right). \textbf{c} Scanning electron microscope image of the Kerr resonator. \textbf{d} Optical microscope image of the two parallel electro-optic phase modulators. In this work, we use only one phase modulator, where the optical waveguide passes through the gaps between the ground-signal coplanar microwave strip lines. \textbf{e} Cross sectional schematic of the Kerr resonator (top) and the electro-optic phase modulator (bottom). \textbf{f} Schematic of the hybrid Kerr-electro-optic frequency comb generation process. A soliton frequency comb (blue) is used as a source, where each soliton comb line generates electro-optic sidebands (red) around it at multiples of the modulation frequency. The  final output of our hybrid comb generator consists of both blue and red lines. \textbf{g} Schematic of the variables defined in our work. $f_{DKS}$: soliton frequency comb spacing, $f_{RF}$: hybrid Kerr-electro-optic frequency comb spacing and electro-optic modulation frequency, $\Delta f$: difference frequency defined by $\Delta f=f_{DKS}-N\cdot f_{RF}$.}
\end{figure*}

\begin{figure*}[t!]
\centering
\includegraphics[width=\linewidth]{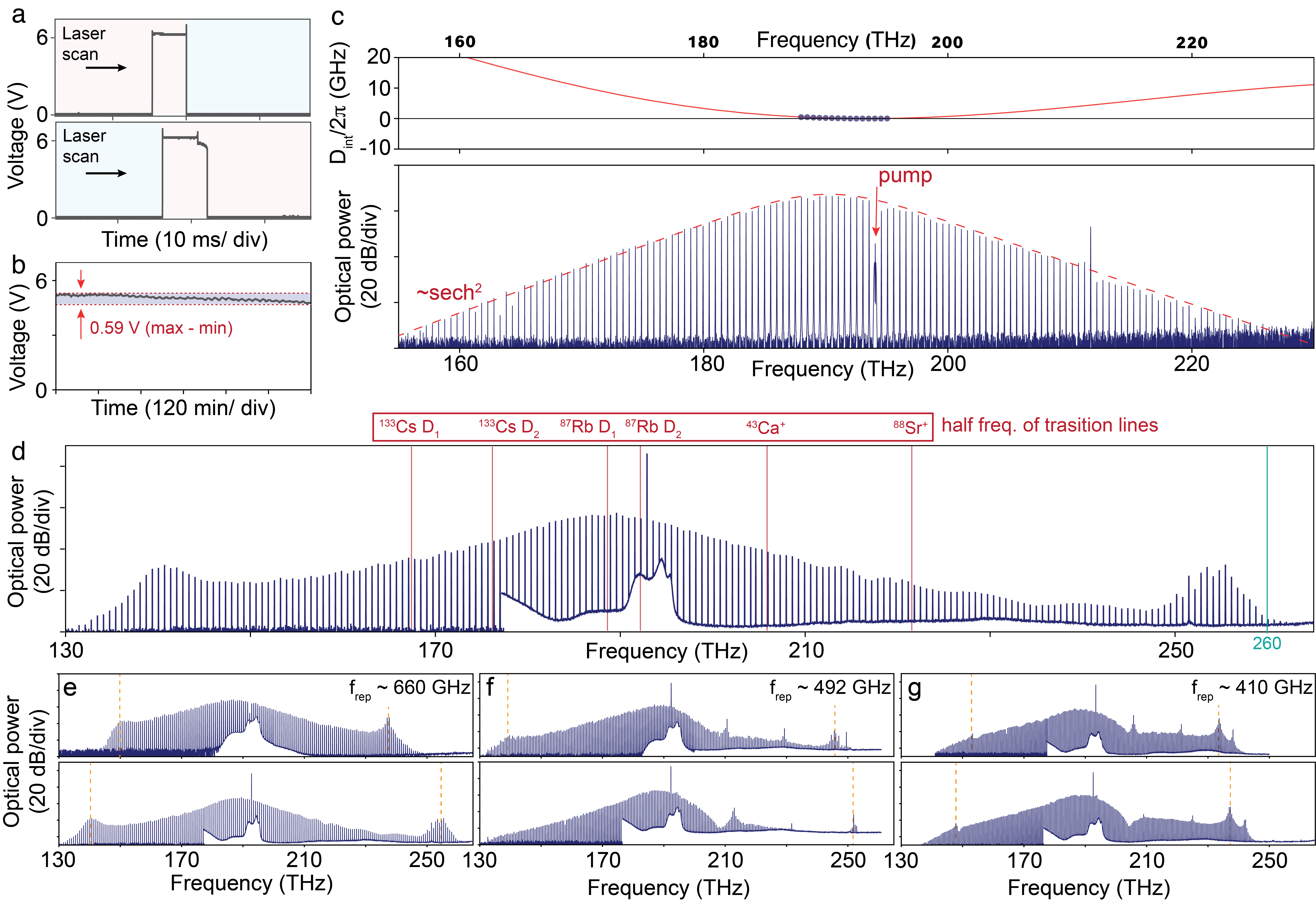}
\caption{\label{fig:Fig2}\textbf{Dissipative Kerr soliton on thin-film lithium niobate.} \textbf{a} Total comb power traces as pump laser scans across a microresonator resonance from the red to blue (top) and blue to red (bottom) directions, accessing flat soliton steps in both directions. The step around 5.4 V is a single-soliton state. \textbf{b} Free-running stability of the single-soliton state in \textbf{a} once manually initiated. The soliton power detected as a voltage remains steady over fifteen hours without any feedback mechanism and the smooth decrease by 0.59 V is dominated by the slowly drifting lensed-fiber coupling into and off of the chip. \textbf{c} Simulated (red curve) and measured (blue dots) integrated dispersion of the microresonator (top) and the single-soliton spectrum (bottom) showcasing a prototypical $\text{sech}^2$ envelope. Simulated dispersion aligns well with our measured results $D_{int}=\omega_{\mu}-\omega_{0}-D_{1}\mu$, where $\mu$ indicates the azimuthal mode index and $\omega_{0}/2\pi$ = 194 THz. The strong pump at approximately 194 THz is filtered out using a fiber-Bragg-grating notch filter. The amplified spontaneous emission noise associated with pump amplification is filtered out using a tunable bandpass filter. \textbf{d} Octave spanning single-soliton spectrum covering 131.3 to 263.2 THz end-to-end, with a spacing of about 660 GHz. Half-frequencies of six atomic transition lines are overlaid (red lines), indicating sufficient spectral coverage over stable atomic transitions for comb stabilization. \textbf{e}-\textbf{g} Dispersion engineered 30, 40, and 50 $\mu$m-radius microresonators, corresponding to spacing between comb lines ($f_{DKS}$) of about 660, 492, and 410 GHz, respectively. The microresonator waveguide width is decreased from top to bottom panels and, in response, the dispersive wave locations (red dashed lines) move away from the pump frequencies, indicating increased anomalous dispersion at the pump frequencies.}
\end{figure*}

\begin{figure*}[t!]
\centering
\includegraphics[width=\linewidth]{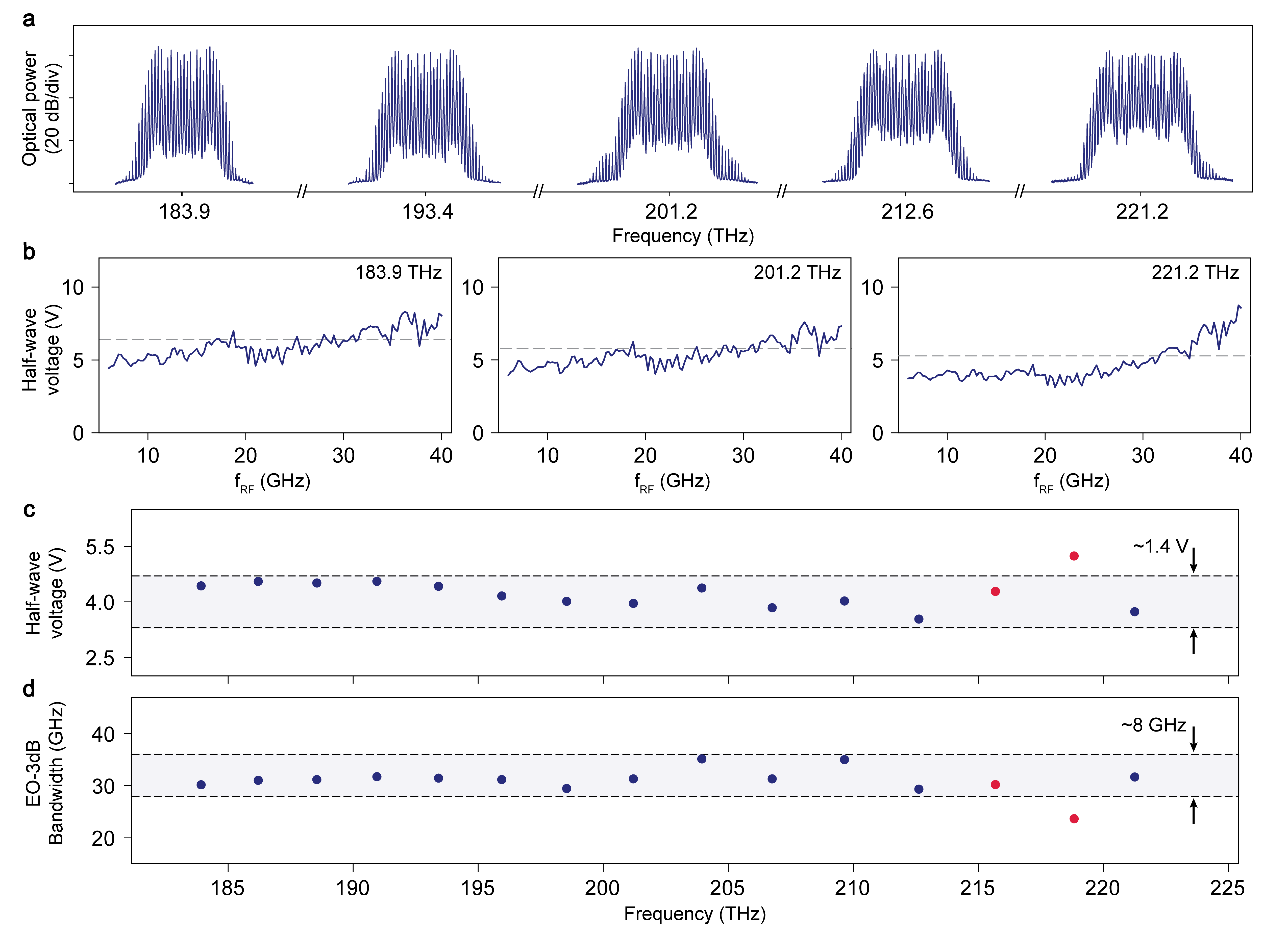}
\caption{\label{fig:Fig3}\textbf{Integrated electro-optic modulation on thin-film lithium niobate.} \textbf{a} Electro-optic frequency combs generated by a 2 cm-long, integrated electro-optic phase modulator at pump frequencies of 183.9, 193.4, 201.2, 212.6, and 221.2 THz (equivalent wavelengths of 1630, 1550, 1490, 1410, and 1355 nm). The modulation frequency is 29.158 GHz, and the electrical power is estimated to be 4.47 W (36.5 dBm). \textbf{b} Half-wave voltage ($V_\pi$) at different modulation frequencies (blue curve) and the $\sqrt{2}$-increased $V_\pi$ (corresponding modulation frequency is the 3 dB electro-optic bandwidth) referenced to the 6 GHz $V_\pi$ (gray dashed line), at three select pump wavelengths. \textbf{c} Pump wavelength dependence of the 6 GHz $V_\pi$ and \textbf{d} the 3 dB electro-optic bandwidth. The 6 GHz $V_\pi$ and 3 dB electro-optic bandwidth variation (blue shaded region enclosed by black dashed lines) over ~30 THz optical bandwidth is about 1.4 V and 8 GHz, respectively. Points near 215 and 219 THz (red dots) are affected by mode-crossings and have relatively higher 6 GHz $V_\pi$ and lower 3 dB electro-optic bandwidth.}
\end{figure*}


\section{Introduction}

\color{black} 
The generation and stabilization of optical and microwave frequencies have had a significant impact on scientific and technological advancement. Optical atomic clocks\cite{ludlow2015optical}, ultrafast lasers\cite{keller2003recent,dudley2006supercontinuum}, precision spectroscopy\cite{coddington2016dual,picque2019frequency}, and advanced communications\cite{marin2017microresonator, rizzo2023massively, Oxenlowe:2022:NaturePhotonics} have been developed based on this foundational technology. Recently, the demand for frequency synthesis and stabilization in both domains has intensified, accompanied by stringent constraints on their scalability and system operational budgets. Chip-scale integration, based on low-loss and highly nonlinear nanophotonic devices, has shown potential to reduce power consumption and system volume by orders of magnitude\cite{spencer2018optical,xiang20233d,kudelin2023photonic,sun2023integrated}. Critically relying on chip-scale frequency comb sources\cite{diddams2020optical,kippenberg2018dissipative,chang2022integrated}, prior demonstrations showcased frequency division of optical carriers down to microwaves\cite{fortier2011generation,xie2017photonic,li2014electro,liu2020photonic,tetsumoto2021optically,kudelin2023photonic,sun2023integrated}, and in reverse, optical frequency synthesis with unparalleled frequency stabilities\cite{spencer2018optical}. Continuous improvement of integrated comb sources foresees major enhancements to such demonstrations, which may significantly benefit from microwave-rate comb spacings coupled with a broad comb span: specifically, microwave-rate spacings of the frequency comb directly link fast-oscillating optical frequencies to electronically detectable microwaves, while a broad comb span enables simultaneous synthesis of distant optical frequencies and their mutual coherence. Given this need, current integrated frequency comb sources are limited by a spacing-span trade-off\cite{kippenberg2018dissipative,chang2022integrated}. Most of such sources to date are either microresonator-based dissipative Kerr solitons (DKS) or electro-optic (EO) frequency combs\cite{kippenberg2018dissipative,boes2023lithium}. While the DKS features octave-spanning bandwidths, the tradeoff between nonlinear enhancement and comb spacing dictates that comb lines are separated by hundreds of GHz or THz frequencies\cite{spencer2018optical,li2017stably,pfeiffer2017octave,liu2021aluminum,weng2021directly}. On the other hand, the EO frequency comb has spacings set by their microwave-rate EO modulation frequencies, but achievable comb spans are limited by the maximally attainable EO interaction strength, resonator mode crossings, and waveguide dispersion\cite{zhang2019broadband,rueda2019resonant,hu2022high,yu2021femtosecond}.

To overcome these limitations, a combination of EO and Kerr nonlinearities have been utilized. Two leading approaches involve EO pulse-pumped DKS\cite{anderson2022zero,cheng2023chip} and cascading Kerr combs with EO modulation\cite{del2012hybrid,drake2019terahertz,moille2023kerr}, but their chip-scale integration has remained elusive. In the first approach, discrete EO modulators, used in conjunction with dispersion compensating fiber, generate high peak power pulses. These pulses are then used to pump near-zero dispersion Kerr resonators, resulting in broadband combs with microwave-separated lines. The second approach involves feeding Kerr combs into bulk EO modulators, which divides the large spacings down to the microwave level. While both approaches have successfully achieved microwave-rate comb spacings and large comb spans, full system integration of these approaches lacks device elements for high peak-power pulse generation, low-loss pulse compression, and efficient EO modulation. Beyond these hybrid approaches, integrated frequency comb generation leveraging the strong Kerr and EO nonlinearities of thin-film lithium niobate (TFLN) was explored, including a DKS electronically referenced by a microresonator EO comb\cite{gong2022monolithic}, EO-tunable DKS\cite{he2023high}, and a single-resonator EO-Raman comb\cite{hu2022high}, but all still face major bandwidth limitations. Therefore, the generation of comb spectra, with simultaneously large span and low spacing, remains a critical hurdle for the field of chip-based frequency combs.

Here, we demonstrate a hybrid Kerr-EO frequency comb with microwave-rate spacing (29.308 GHz) and broad bandwidth (75.9 THz). In our approach, the intermodal Kerr interaction in a microresonator produces a mode-locked, broadband, and near THz-rate DKS frequency comb, while subsequent non-resonant EO division is used to coherently densify the source spacing down to microwave frequencies over the source bandwidth. Importantly, we demonstrate an octave-spanning DKS on TFLN, creating a prospective avenue for self-referenced integrated frequency combs utilizing EO \cite{wang2018integrated,xu2022dual} and second harmonic generation\cite{wang2018ultrahigh,lu2019periodically,mckenna2022ultra} on the same platform\cite{zhu2021integrated}. The large Kerr and EO coefficients in TFLN relax optical and electrical requirements compared to previous hybrid-comb demonstrations and render our approach as a clear path forward towards broadband, microwave-rate optical frequency combs from a single photonic material.

\section{Results}
\vspace{1ex}
\noindent\textbf{Device design and experimental approach}\\
Our hybrid Kerr-EO approach for frequency comb generation on TFLN cascades a dispersion-engineered microresonator and a high-speed, efficient EO phase modulator. The microresonator outputs a DKS comb with a comb spacing in the range of hundreds of GHz. Each soliton line is subsequently phase modulated, and EO sidebands are formed in a cascaded manner around each DKS comb line such that the original DKS comb spacing is fully divided into microwave-rate spacing set by the EO modulation frequency. The required integrated-photonic components are schematically shown in Fig. 1\textbf{a}. Taking this approach, we developed Kerr microresonators and EO phase modulators on Z-cut and X-cut TFLN chips, respectively. Images of the fabricated Kerr microresonators and EO phase modulator are shown in Fig. 1\textbf{b}-\textbf{e}. The working principle of the hybrid Kerr-EO approach and final frequency comb output is schematically illustrated in Fig. 1\textbf{f}, \textbf{g}.

\vspace{1ex}
\noindent\textbf{Octave-spanning DKS on TFLN}\\
The TFLN photonic platform is a prime candidate on which to implement the hybrid Kerr-EO approach for broadband, microwave-rate frequency comb generation, due to its large Kerr and EO nonlinearities. It hosts low-loss microresonators\cite{zhang2017monolithic} supporting broadband DKS states initiated manually\cite{he2019self}, which also exhibit excellent free-running stability\cite{wan2023photorefraction}, owing to the material’s photorefractive-induced, pump-resonance self-locking when the pump is red-detuned. Much like the thermo-optic self-stability of microresonators\cite{carmon2004dynamical} when driven by a blue-detuned pump, common to nearly all other photonic platforms, the TFLN microresonator resonances stabilize against fluctuations in the pump frequency when the pump is on the red-side of resonance, which coincides with the region of DKS existence. This phenomenon differs from DKS generation in other platforms, where thermo-optic bistability under a red-detuned pump hinders stabilization into a DKS state\cite{herr2014temporal}. The self-starting nature of DKSs on TFLN simplifies the system complexity involved in their formation and stabilization. In Fig. 2\textbf{a}, we show that for a suitable resonance, when the pump frequency slowly sweeps across resonance starting from the red or blue detuned side, flat steps indicative of DKS states are present, irrespective of the initial condition. Once a DKS state emerges it is self-stable, as evidenced in Fig. 2\textbf{b} by the finite DKS power measured over 15 hours without any feedback acting on the microresonator or the pump laser. We show a prototypical single DKS state generated in the fundamental transverse-electric mode in Fig. 2\textbf{c}, where a strong quadratic term dominates the total microresonator dispersion $D_{\text{int}}$, giving rise to a $\text{sech}^2$ spectrum under a moderate on-chip pump power of 189 mW.

To further unlock the potential of DKSs in TFLN, we demonstrate an octave-spanning single soliton frequency comb from 131.3 THz to 263.8 THz with a spacing of 660 GHz, initiated under 372 mW of on-chip pump power (Fig. 2\textbf{d}). This is enabled by dispersion engineering via precisely tuning the radius, waveguide width, and waveguide height (etch depth) of the TFLN microresonators: for the three ring radii chosen (30, 40, and 50 $\mu$m in Fig. 2\textbf{e}-\textbf{g}, respectively), we vary the microresonator waveguide width (decreasing width from top to bottom panels, in each of Fig. 2\textbf{e}-\textbf{g}) while fixing the waveguide height to about 345 nm. Using on-chip pump powers between 100 and 250 mW, we consistently generate well-controlled DKS spectra broadened by dual dispersive waves\cite{brasch2016photonic,li2017stably,spencer2018optical} into frequencies of normal dispersion, significantly increasing the comb spans. As the waveguide width is decreased, the total comb span broadens as the dispersive wave locations move away from the pump, consistent with expectation (Methods). The broadband DKSs, initiated under moderate on-chip pump powers, serve as ideal sources for practical hybrid Kerr-EO combs.

\begin{figure*}[t!]
\centering
\includegraphics[width=\linewidth]{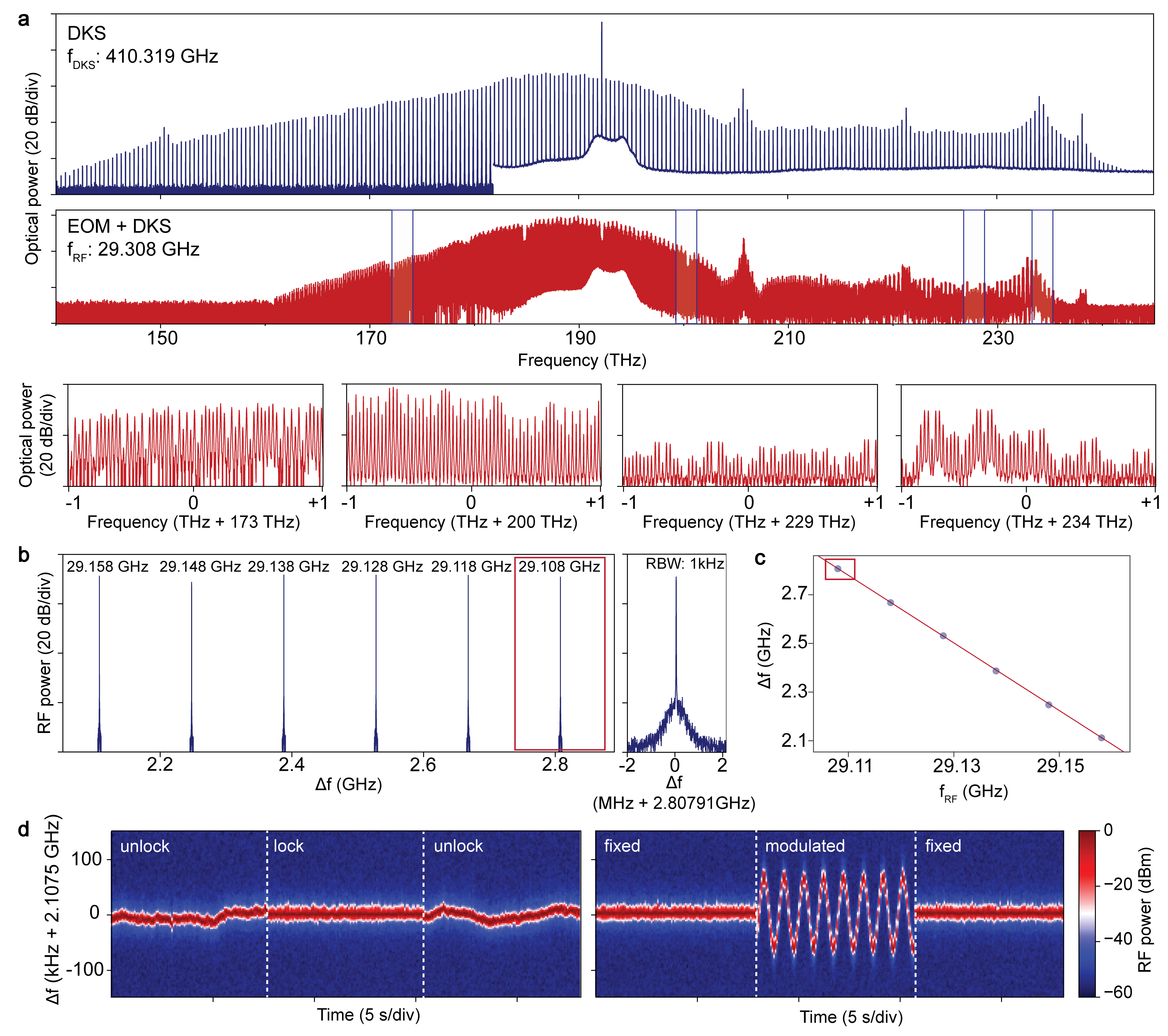}
\caption{\label{fig:Fig4}\textbf{Hybrid Kerr-electro-optic frequency comb on thin-film lithium niobate.} \textbf{a} Single-soliton spectrum (blue, top) with a spacing ($f_{DKS}$) of 410.319 GHz between comb lines and electro-optically modulated single-soliton spectrum (red, bottom) with a 14-times divided spacing of 29.308 GHz. The latter spectrum is zoomed-in over 2 THz windows centered around 173, 200, 229, and 234 THz. \textbf{b} Nonzero difference frequencies ($\Delta f$), detected as an electro-optic beatnote, as the electro-optic modulation frequency is stepped from 29.108 GHz to 29.158 GHz in intervals of 0.01 GHz. When the modulation frequency is 29.108 GHz, a zoom-in of the beatnote over a 4 MHz window centered around 2.80791 GHz shows a narrow linewidth under 1 kHz of resolution bandwidth. \textbf{c} Linear fit of the beatnote frequencies to the modulation frequencies in \textbf{b}. The absolute value of the slope is $N=14$, the number of electro-optic sidebands required to completely divide the single soliton spacing. \textbf{d} Spectrogram of $\Delta f$ when the modulation frequency is fixed at 29.158 GHz. Left panel shows the progression of $\Delta f$ when it is unlocked, phase locked to a microwave reference oscillator, and unlocked again. Right panel shows the progression of $\Delta f$ when it is phase locked to a microwave reference oscillator that is fixed frequency, sinusoidally modulated in frequency, and fixed frequency again.}
\end{figure*}

\vspace{1ex}
\noindent\textbf{High-speed EO modulation on TFLN}\\
Next, we characterize the second piece to our approach, a TFLN EO phase modulator operated at multiple wavelengths sampled within the source DKS bandwidth. The waveguide-based integrated modulator operates with a travelling-wave electrode configuration. 2 cm-long microwave strip lines are used to induce a larger optical phase shift and reduce the V$_{\pi}$ value, simultaneously exhibiting large operational bandwidth\cite{wang2018integrated}. Single-tone microwaves at frequency 29.158 GHz are continuously delivered to the EO phase modulator electrodes. It is important to note that the waveguide-based modulator operates with a broad, uninterrupted microwave band, over a continuous optical frequency span, as opposed to the cavity-based modulators\cite{gong2022monolithic}. Additionally, our broadband modulator can significantly ease the complexity involved in f-2f stabilization on a chip\cite{spencer2018optical} -- the carrier offset frequency, often larger than the detector bandwidth, can be divided down using the modulator. Fig. 3 presents a broadband operation of our modulator at both microwave and optical frequencies.

Five continuous-wave optical frequencies in the telecommunications L, C, S, and E bands are separately coupled onto the chip through bilayer-taper mode converters, and EO sidebands are generated around each frequency, as shown in Fig. 3\textbf{a}. At all wavelengths, the microwave electrical power is estimated to be 4.47 W (36.5 dBm), showcasing the excellent power-handling capability of our modulator. We characterized the modulator half-wave voltage ($V_\pi$) and EO bandwidth at these wavelengths by measuring the modulation-frequency dependent $V_\pi$ (Methods), and the results are shown in Fig. 3\textbf{b}-\textbf{d}. While the $V_\pi$ and EO bandwidth show slight dependencies on the optical wavelength, likely due to varying optical-microwave spatial mode overlaps and dispersion-induced velocity mismatches\cite{wang2018integrated}, they are sufficient considering the total bandwidth of our DKS combs and the required number of EO sidebands for complete EO division of their spacings. 

\vspace{1ex}
\noindent\textbf{Hybrid Kerr-EO integrated frequency comb}\\
Combining self-starting and stable TFLN DKSs with efficient and high-speed EO modulation, a fully integrated hybrid Kerr-EO comb is realizable. By sending a low-threshold, 410.319 GHz-spaced DKS into our EO phase modulator, we obtain a hybrid Kerr-EO comb spanning 75.9 THz end-to-end with a spacing of 29.308 GHz (Fig. 4\textbf{a}). The relationship between the spacings is given by $f_{DKS}=N\cdot f_{RF}+\Delta f$, where $f_{DKS}$ and $f_{RF}$ are the DKS and hybrid Kerr-EO comb spacings, $N$ is the number of EO sidebands to fully divide one DKS spacing, and $\Delta f$ is the difference frequency from perfect EO division. These quantities are also schematically illustrated in Fig. 1\textbf{g}. Note that $N$ is the largest integer for which $|\Delta f|<f_{RF}$ holds. When the EO sidebands perfectly divide the DKS spacing, $f_{DKS}/f_{RF} =N$ and $\Delta f=0$. The DKS spectrum is filled in with microwave-separated EO lines, as shown in the zoom-in windows of Fig. 4\textbf{a}. Power variation in the comb lines within each window comes from the EO sideband envelope determined by the local modulation depth, and that between windows comes from the frequency-dependent $V_\pi$. The optical pump power for DKS generation is 125 mW (21 dBm) and the electrical power to drive the EO phase modulator is 2.51 W (34.5 dBm), both on-chip. The excellent connectivity of our comb offers densely positioned optical frequency references over its entire bandwidth. The free-running uptime of this broadband reference is determined by that of the source DKS, which we have shown, in Fig. 2\textbf{b}, can be more than 15 hours even in an uncontrolled environment. It is also important to note that our setup didn't incorporate any external optical amplifiers or dispersion compensating elements between the DKS and EOM chips.

Next, we set $\Delta f$ to be a nonzero value of around 2.107 GHz to elucidate the spacing relationship. As the microwave drive frequency (which sets $f_{RF}$) is linearly decreased in steps of 0.01 GHz, the $\Delta f$ linearly increases, as shown in Fig. 4\textbf{b}. Fitting $f_{RF}$ against $\Delta f$, the slope gives precisely $N=14$ in support of the spacing relationship. Confirming this relationship, our single integrated EO phase modulator thus demonstrated a direct measurement of the near THz spacing. Importantly, this measurement via efficient EO division transduces the near THz spacing $f_{DKS}$ into a GHz difference frequency $\Delta f$ compatible with detection via fast electronics. Further, the detected $\Delta f$ may then be compared and phase locked to a stable microwave reference by feeding back onto the DKS pump frequency, which in turn stabilizes the original DKS spacing. In Fig. 4\textbf{d}, we phase lock the $\Delta f$ around 2.107 GHz to 32 times a microwave reference oscillator set at frequency 65.865 MHz. When unlocked, the $\Delta f$ fluctuates according to the fluctuation in $f_{DKS}$. When the lock is engaged, $\Delta f$ is pinned to a center frequency of 2.107 GHz with a 30 kHz linewidth. Sinusoidally modulating the reference oscillator frequency during the phase lock, the $\Delta f$ follows the reference modulation and so does $f_{DKS}$, indicating detection-based, integrated EO control over the near THz-rate DKS spacing. Such a method for DKS spacing control stands in contrast to EO-based parametric seeding techniques which may only control microwave-rate, span-limited Kerr combs\cite{he2023high,papp2013parametric} or, in principle, seed THz combs with very poor duty cycles. Thus, our hybrid Kerr-EO approach towards TFLN-based frequency combs not only generates broadband, microwave-rate combs, but also opens a route towards on-chip methods for the detection, stabilization, and control over large DKS spacings located in previously inaccessible THz regimes.

\section{Conclusion and outlook}
In summary, we outlined a hybrid Kerr-EO approach for integrated frequency comb generation and demonstrated TFLN as a material platform that realizes this approach completely. Our work is enabled by the outstanding second and third-order nonlinear properties of TFLN, hosting self-starting, low threshold, and stable free-running THz-rate DKSs, while supporting on-chip EO phase modulation capabilities with excellent bandwidth, efficiency, and speed. Currently, the comb line power difference between the DKS source and the hybrid Kerr-EO comb is attributed to EO power redistribution into the sidebands (about 10 dB pump mode suppression near 1550 nm),  the off-chip coupling loss from the DKS chip (about 6 dB), and the insertion loss of the EO modulator chip (about 8 dB, both chip facets and propagation loss included). The latter two losses, totaling 14 dB, does not present a fundamental limitation and can be reduced significantly using improved coupling methods such as hybrid integration taking advantage of low-loss, multi-layer couplers\cite{liu2022ultra} or monolithic integration of TFLN crystal cuts\cite{zhou2023monolithically}. The excess fiber length between the chips (more than 10 meters) led to attenuation of longer wavelength components due to silica absorption, a source of loss which also may be eliminated by advanced coupling methods. We note that the device stack of both DKS and EO phase modulator chips are sufficiently similar for the seamless implementation of these methods. Further, the electrical power requirement of driving the EO phase modulator may be significantly reduced down to hundreds of milliwatts, employing state-of-the-art modulators (Methods)\cite{xu2022dual,zhu2022spectral}.

Our integrated comb source, serving as a broadband, microwave-rate frequency reference spanning 75.9 THz around the telecommunications band, is suitable for the measurement, control, and mutual stabilization of optical frequencies. Such frequencies may take the form of independent chip-scale lasers\cite{jin2021hertz,de2021iii,shams2022electrically,li2022integrated,guo2023ultrafast,snigirev2023ultrafast} for gap-free and frequency-accurate laser spectroscopy\cite{del2009frequency,shi2023frequency}. Full stabilization of the hybrid Kerr-EO reference, such that the carrier-envelope offset frequency and comb spacing are simultaneously locked, can be facilitated through Kerr-induced synchronization\cite{moille2023kerr}, interferometric self-referencing\cite{spencer2018optical,brasch2017self}, or electronic feedback locking of a comb line to a stable atomic transition\cite{papp2014microresonator,newman2019architecture}. Notably, in many applications, a fully connected comb over the entire DKS span may not be necessary and a locked THz-rate DKS spacing is sufficient. The demonstrated locking and modulation of the difference frequency $\Delta f$ of our hybrid Kerr-EO comb enables direct stabilization and control of the original, large DKS spacing, which may form the basis for THz-related applications such as radar and high-bandwidth wireless communications\cite{koenig2013wireless,wang2021towards}. Further, a hybrid Kerr-EO comb may extend conventional optical\cite{fortier2011generation,xie2017photonic,liu2020photonic,tetsumoto2021optically,kudelin2023photonic,sun2023integrated} and EO\cite{li2014electro} frequency division schemes towards integrated synthesizers of ultra-stable microwaves that promise orders of magnitude improvement in division factor when a cascaded DKS-EO frequency division is considered. Therefore, our integrated hybrid Kerr-EO frequency comb may lay the foundation for a variety of next-generation chip-scale devices that play a key role in system-level applications\cite{marin2017microresonator,Bhaskaran:2021:Nature,Moss:2021:Nature,spencer2018optical,kudelin2023photonic,sun2023integrated} of integrated photonics, where the power of optical frequency combs\cite{diddams2020optical} is paramount.

\bibliography{Reference}

\begin{thebibliography}{10}
\expandafter\ifx\csname url\endcsname\relax
  \def\url#1{\texttt{#1}}\fi
\expandafter\ifx\csname urlprefix\endcsname\relax\def\urlprefix{URL }\fi
\providecommand{\bibinfo}[2]{#2}
\providecommand{\eprint}[2][]{\url{#2}}

\bibitem{udem2002optical}
\bibinfo{author}{Udem, T.}, \bibinfo{author}{Holzwarth, R.} \& \bibinfo{author}{H{\"a}nsch, T.~W.}
\newblock \bibinfo{title}{Optical frequency metrology}.
\newblock \emph{\bibinfo{journal}{Nature}} \textbf{\bibinfo{volume}{416}}, \bibinfo{pages}{233--237} (\bibinfo{year}{2002}).

\bibitem{diddams2020optical}
\bibinfo{author}{Diddams, S.~A.}, \bibinfo{author}{Vahala, K.} \& \bibinfo{author}{Udem, T.}
\newblock \bibinfo{title}{Optical frequency combs: Coherently uniting the electromagnetic spectrum}.
\newblock \emph{\bibinfo{journal}{Science}} \textbf{\bibinfo{volume}{369}}, \bibinfo{pages}{eaay3676} (\bibinfo{year}{2020}).

\bibitem{suh2016microresonator}
\bibinfo{author}{Suh, M.-G.}, \bibinfo{author}{Yang, Q.-F.}, \bibinfo{author}{Yang, K.~Y.}, \bibinfo{author}{Yi, X.} \& \bibinfo{author}{Vahala, K.~J.}
\newblock \bibinfo{title}{Microresonator soliton dual-comb spectroscopy}.
\newblock \emph{\bibinfo{journal}{Science}} \textbf{\bibinfo{volume}{354}}, \bibinfo{pages}{600--603} (\bibinfo{year}{2016}).

\bibitem{coddington2016dual}
\bibinfo{author}{Coddington, I.}, \bibinfo{author}{Newbury, N.} \& \bibinfo{author}{Swann, W.}
\newblock \bibinfo{title}{Dual-comb spectroscopy}.
\newblock \emph{\bibinfo{journal}{Optica}} \textbf{\bibinfo{volume}{3}}, \bibinfo{pages}{414--426} (\bibinfo{year}{2016}).

\bibitem{picque2019frequency}
\bibinfo{author}{Picqu{\'e}, N.} \& \bibinfo{author}{H{\"a}nsch, T.~W.}
\newblock \bibinfo{title}{Frequency comb spectroscopy}.
\newblock \emph{\bibinfo{journal}{Nature Photonics}} \textbf{\bibinfo{volume}{13}}, \bibinfo{pages}{146--157} (\bibinfo{year}{2019}).

\bibitem{spencer2018optical}
\bibinfo{author}{Spencer, D.~T.} \emph{et~al.}
\newblock \bibinfo{title}{An optical-frequency synthesizer using integrated photonics}.
\newblock \emph{\bibinfo{journal}{Nature}} \textbf{\bibinfo{volume}{557}}, \bibinfo{pages}{81--85} (\bibinfo{year}{2018}).

\bibitem{fortier2011generation}
\bibinfo{author}{Fortier, T.~M.} \emph{et~al.}
\newblock \bibinfo{title}{Generation of ultrastable microwaves via optical frequency division}.
\newblock \emph{\bibinfo{journal}{Nature Photonics}} \textbf{\bibinfo{volume}{5}}, \bibinfo{pages}{425--429} (\bibinfo{year}{2011}).

\bibitem{li2014electro}
\bibinfo{author}{Li, J.}, \bibinfo{author}{Yi, X.}, \bibinfo{author}{Lee, H.}, \bibinfo{author}{Diddams, S.~A.} \& \bibinfo{author}{Vahala, K.~J.}
\newblock \bibinfo{title}{Electro-optical frequency division and stable microwave synthesis}.
\newblock \emph{\bibinfo{journal}{Science}} \textbf{\bibinfo{volume}{345}}, \bibinfo{pages}{309--313} (\bibinfo{year}{2014}).

\bibitem{xie2017photonic}
\bibinfo{author}{Xie, X.} \emph{et~al.}
\newblock \bibinfo{title}{Photonic microwave signals with zeptosecond-level absolute timing noise}.
\newblock \emph{\bibinfo{journal}{Nature photonics}} \textbf{\bibinfo{volume}{11}}, \bibinfo{pages}{44--47} (\bibinfo{year}{2017}).

\bibitem{liu2020photonic}
\bibinfo{author}{Liu, J.} \emph{et~al.}
\newblock \bibinfo{title}{Photonic microwave generation in the x-and k-band using integrated soliton microcombs}.
\newblock \emph{\bibinfo{journal}{Nature Photonics}} \textbf{\bibinfo{volume}{14}}, \bibinfo{pages}{486--491} (\bibinfo{year}{2020}).

\bibitem{tetsumoto2021optically}
\bibinfo{author}{Tetsumoto, T.} \emph{et~al.}
\newblock \bibinfo{title}{Optically referenced 300 ghz millimetre-wave oscillator}.
\newblock \emph{\bibinfo{journal}{Nature Photonics}} \textbf{\bibinfo{volume}{15}}, \bibinfo{pages}{516--522} (\bibinfo{year}{2021}).

\bibitem{papp2014microresonator}
\bibinfo{author}{Papp, S.~B.} \emph{et~al.}
\newblock \bibinfo{title}{Microresonator frequency comb optical clock}.
\newblock \emph{\bibinfo{journal}{Optica}} \textbf{\bibinfo{volume}{1}}, \bibinfo{pages}{10--14} (\bibinfo{year}{2014}).

\bibitem{newman2019architecture}
\bibinfo{author}{Newman, Z.~L.} \emph{et~al.}
\newblock \bibinfo{title}{Architecture for the photonic integration of an optical atomic clock}.
\newblock \emph{\bibinfo{journal}{Optica}} \textbf{\bibinfo{volume}{6}}, \bibinfo{pages}{680--685} (\bibinfo{year}{2019}).

\bibitem{herr2014temporal}
\bibinfo{author}{Herr, T.} \emph{et~al.}
\newblock \bibinfo{title}{Temporal solitons in optical microresonators}.
\newblock \emph{\bibinfo{journal}{Nature Photonics}} \textbf{\bibinfo{volume}{8}}, \bibinfo{pages}{145--152} (\bibinfo{year}{2014}).

\bibitem{kippenberg2018dissipative}
\bibinfo{author}{Kippenberg, T.~J.}, \bibinfo{author}{Gaeta, A.~L.}, \bibinfo{author}{Lipson, M.} \& \bibinfo{author}{Gorodetsky, M.~L.}
\newblock \bibinfo{title}{Dissipative kerr solitons in optical microresonators}.
\newblock \emph{\bibinfo{journal}{Science}} \textbf{\bibinfo{volume}{361}}, \bibinfo{pages}{eaan8083} (\bibinfo{year}{2018}).

\bibitem{zhang2019broadband}
\bibinfo{author}{Zhang, M.} \emph{et~al.}
\newblock \bibinfo{title}{Broadband electro-optic frequency comb generation in a lithium niobate microring resonator}.
\newblock \emph{\bibinfo{journal}{Nature}} \textbf{\bibinfo{volume}{568}}, \bibinfo{pages}{373--377} (\bibinfo{year}{2019}).

\bibitem{rueda2019resonant}
\bibinfo{author}{Rueda, A.}, \bibinfo{author}{Sedlmeir, F.}, \bibinfo{author}{Kumari, M.}, \bibinfo{author}{Leuchs, G.} \& \bibinfo{author}{Schwefel, H.~G.}
\newblock \bibinfo{title}{Resonant electro-optic frequency comb}.
\newblock \emph{\bibinfo{journal}{Nature}} \textbf{\bibinfo{volume}{568}}, \bibinfo{pages}{378--381} (\bibinfo{year}{2019}).

\bibitem{hu2022high}
\bibinfo{author}{Hu, Y.} \emph{et~al.}
\newblock \bibinfo{title}{High-efficiency and broadband on-chip electro-optic frequency comb generators}.
\newblock \emph{\bibinfo{journal}{Nature Photonics}} \textbf{\bibinfo{volume}{16}}, \bibinfo{pages}{679--685} (\bibinfo{year}{2022}).

\bibitem{yu2021femtosecond}
\bibinfo{author}{Yu, M.} \emph{et~al.}
\newblock \bibinfo{title}{Integrated femtosecond pulse generator on thin-film lithium niobate}.
\newblock \emph{\bibinfo{journal}{Nature}} \textbf{\bibinfo{volume}{612}}, \bibinfo{pages}{252--258} (\bibinfo{year}{2022}).

\bibitem{boes2023lithium}
\bibinfo{author}{Boes, A.} \emph{et~al.}
\newblock \bibinfo{title}{Lithium niobate photonics: Unlocking the electromagnetic spectrum}.
\newblock \emph{\bibinfo{journal}{Science}} \textbf{\bibinfo{volume}{379}}, \bibinfo{pages}{eabj4396} (\bibinfo{year}{2023}).

\bibitem{shi2023frequency}
\bibinfo{author}{Shi, B.} \emph{et~al.}
\newblock \bibinfo{title}{Frequency-comb-linearized, widely tunable lasers for coherent ranging}.
\newblock \emph{\bibinfo{journal}{arXiv preprint arXiv:2308.15875}}  (\bibinfo{year}{2023}).

\bibitem{koenig2013wireless}
\bibinfo{author}{Koenig, S.} \emph{et~al.}
\newblock \bibinfo{title}{Wireless sub-thz communication system with high data rate}.
\newblock \emph{\bibinfo{journal}{Nature Photonics}} \textbf{\bibinfo{volume}{7}}, \bibinfo{pages}{977--981} (\bibinfo{year}{2013}).

\bibitem{wang2021towards}
\bibinfo{author}{Wang, B.} \emph{et~al.}
\newblock \bibinfo{title}{Towards high-power, high-coherence, integrated photonic mmwave platform with microcavity solitons}.
\newblock \emph{\bibinfo{journal}{Light: Science \& Applications}} \textbf{\bibinfo{volume}{10}}, \bibinfo{pages}{4} (\bibinfo{year}{2021}).

\bibitem{marin2017microresonator}
\bibinfo{author}{Marin-Palomo, P.} \emph{et~al.}
\newblock \bibinfo{title}{Microresonator-based solitons for massively parallel coherent optical communications}.
\newblock \emph{\bibinfo{journal}{Nature}} \textbf{\bibinfo{volume}{546}}, \bibinfo{pages}{274--279} (\bibinfo{year}{2017}).

\bibitem{Oxenlowe:2022:NaturePhotonics}
\bibinfo{author}{J{\o}rgensen, A.} \emph{et~al.}
\newblock \bibinfo{title}{Petabit-per-second data transmission using a chip-scale microcomb ring resonator source}.
\newblock \emph{\bibinfo{journal}{Nature Photonics}} \textbf{\bibinfo{volume}{16}}, \bibinfo{pages}{798--802} (\bibinfo{year}{2022}).

\bibitem{Vuckovic:2022:NatureCommun}
\bibinfo{author}{Yang, K.~Y.} \emph{et~al.}
\newblock \bibinfo{title}{Multi-dimensional data transmission using inverse-designed silicon photonics and microcombs}.
\newblock \emph{\bibinfo{journal}{Nature Communications}} \textbf{\bibinfo{volume}{13}}, \bibinfo{pages}{7862} (\bibinfo{year}{2022}).

\bibitem{rizzo2023massively}
\bibinfo{author}{Rizzo, A.} \emph{et~al.}
\newblock \bibinfo{title}{Massively scalable kerr comb-driven silicon photonic link}.
\newblock \emph{\bibinfo{journal}{Nature Photonics}} \textbf{\bibinfo{volume}{17}}, \bibinfo{pages}{781--790} (\bibinfo{year}{2023}).

\bibitem{ludlow2015optical}
\bibinfo{author}{Ludlow, A.~D.}, \bibinfo{author}{Boyd, M.~M.}, \bibinfo{author}{Ye, J.}, \bibinfo{author}{Peik, E.} \& \bibinfo{author}{Schmidt, P.~O.}
\newblock \bibinfo{title}{Optical atomic clocks}.
\newblock \emph{\bibinfo{journal}{Reviews of Modern Physics}} \textbf{\bibinfo{volume}{87}}, \bibinfo{pages}{637} (\bibinfo{year}{2015}).

\bibitem{keller2003recent}
\bibinfo{author}{Keller, U.}
\newblock \bibinfo{title}{Recent developments in compact ultrafast lasers}.
\newblock \emph{\bibinfo{journal}{Nature}} \textbf{\bibinfo{volume}{424}}, \bibinfo{pages}{831--838} (\bibinfo{year}{2003}).

\bibitem{dudley2006supercontinuum}
\bibinfo{author}{Dudley, J.~M.}, \bibinfo{author}{Genty, G.} \& \bibinfo{author}{Coen, S.}
\newblock \bibinfo{title}{Supercontinuum generation in photonic crystal fiber}.
\newblock \emph{\bibinfo{journal}{Reviews of modern physics}} \textbf{\bibinfo{volume}{78}}, \bibinfo{pages}{1135} (\bibinfo{year}{2006}).

\bibitem{xiang20233d}
\bibinfo{author}{Xiang, C.} \emph{et~al.}
\newblock \bibinfo{title}{3d integration enables ultralow-noise isolator-free lasers in silicon photonics}.
\newblock \emph{\bibinfo{journal}{Nature}} \textbf{\bibinfo{volume}{620}}, \bibinfo{pages}{78--85} (\bibinfo{year}{2023}).

\bibitem{kudelin2023photonic}
\bibinfo{author}{Kudelin, I.} \emph{et~al.}
\newblock \bibinfo{title}{Photonic chip-based low noise microwave oscillator}.
\newblock \emph{\bibinfo{journal}{arXiv preprint arXiv:2307.08937}}  (\bibinfo{year}{2023}).

\bibitem{sun2023integrated}
\bibinfo{author}{Sun, S.} \emph{et~al.}
\newblock \bibinfo{title}{Integrated optical frequency division for stable microwave and mmwave generation}.
\newblock \emph{\bibinfo{journal}{arXiv preprint arXiv:2305.13575}}  (\bibinfo{year}{2023}).

\bibitem{chang2022integrated}
\bibinfo{author}{Chang, L.}, \bibinfo{author}{Liu, S.} \& \bibinfo{author}{Bowers, J.~E.}
\newblock \bibinfo{title}{Integrated optical frequency comb technologies}.
\newblock \emph{\bibinfo{journal}{Nature Photonics}} \textbf{\bibinfo{volume}{16}}, \bibinfo{pages}{95--108} (\bibinfo{year}{2022}).

\bibitem{li2017stably}
\bibinfo{author}{Li, Q.} \emph{et~al.}
\newblock \bibinfo{title}{Stably accessing octave-spanning microresonator frequency combs in the soliton regime}.
\newblock \emph{\bibinfo{journal}{Optica}} \textbf{\bibinfo{volume}{4}}, \bibinfo{pages}{193--203} (\bibinfo{year}{2017}).

\bibitem{pfeiffer2017octave}
\bibinfo{author}{Pfeiffer, M.~H.} \emph{et~al.}
\newblock \bibinfo{title}{Octave-spanning dissipative kerr soliton frequency combs in ${Si}_{3}{N}_{4}$ microresonators}.
\newblock \emph{\bibinfo{journal}{Optica}} \textbf{\bibinfo{volume}{4}}, \bibinfo{pages}{684--691} (\bibinfo{year}{2017}).

\bibitem{liu2021aluminum}
\bibinfo{author}{Liu, X.} \emph{et~al.}
\newblock \bibinfo{title}{Aluminum nitride nanophotonics for beyond-octave soliton microcomb generation and self-referencing}.
\newblock \emph{\bibinfo{journal}{Nature Communications}} \textbf{\bibinfo{volume}{12}}, \bibinfo{pages}{5428} (\bibinfo{year}{2021}).

\bibitem{weng2021directly}
\bibinfo{author}{Weng, H.} \emph{et~al.}
\newblock \bibinfo{title}{Directly accessing octave-spanning dissipative kerr soliton frequency combs in an aln microresonator}.
\newblock \emph{\bibinfo{journal}{Photonics Research}} \textbf{\bibinfo{volume}{9}}, \bibinfo{pages}{1351--1357} (\bibinfo{year}{2021}).

\bibitem{anderson2022zero}
\bibinfo{author}{Anderson, M.~H.} \emph{et~al.}
\newblock \bibinfo{title}{Zero dispersion kerr solitons in optical microresonators}.
\newblock \emph{\bibinfo{journal}{Nature communications}} \textbf{\bibinfo{volume}{13}}, \bibinfo{pages}{4764} (\bibinfo{year}{2022}).

\bibitem{cheng2023chip}
\bibinfo{author}{Cheng, R.} \emph{et~al.}
\newblock \bibinfo{title}{On-chip synchronous pumped $\chi^{(3)}$ optical parametric oscillator on thin-film lithium niobate}.
\newblock \emph{\bibinfo{journal}{arXiv preprint arXiv:2304.12878}}  (\bibinfo{year}{2023}).

\bibitem{del2012hybrid}
\bibinfo{author}{Del’Haye, P.}, \bibinfo{author}{Papp, S.~B.} \& \bibinfo{author}{Diddams, S.~A.}
\newblock \bibinfo{title}{Hybrid electro-optically modulated microcombs}.
\newblock \emph{\bibinfo{journal}{Physical review letters}} \textbf{\bibinfo{volume}{109}}, \bibinfo{pages}{263901} (\bibinfo{year}{2012}).

\bibitem{drake2019terahertz}
\bibinfo{author}{Drake, T.~E.} \emph{et~al.}
\newblock \bibinfo{title}{Terahertz-rate kerr-microresonator optical clockwork}.
\newblock \emph{\bibinfo{journal}{Physical Review X}} \textbf{\bibinfo{volume}{9}}, \bibinfo{pages}{031023} (\bibinfo{year}{2019}).

\bibitem{moille2023kerr}
\bibinfo{author}{Moille, G.} \emph{et~al.}
\newblock \bibinfo{title}{Kerr-induced synchronization of a cavity soliton to an optical reference}.
\newblock \emph{\bibinfo{journal}{Nature}} \textbf{\bibinfo{volume}{624}}, \bibinfo{pages}{267--274} (\bibinfo{year}{2023}).

\bibitem{gong2022monolithic}
\bibinfo{author}{Gong, Z.}, \bibinfo{author}{Shen, M.}, \bibinfo{author}{Lu, J.}, \bibinfo{author}{Surya, J.~B.} \& \bibinfo{author}{Tang, H.~X.}
\newblock \bibinfo{title}{Monolithic kerr and electro-optic hybrid microcombs}.
\newblock \emph{\bibinfo{journal}{Optica}} \textbf{\bibinfo{volume}{9}}, \bibinfo{pages}{1060--1065} (\bibinfo{year}{2022}).

\bibitem{he2023high}
\bibinfo{author}{He, Y.} \emph{et~al.}
\newblock \bibinfo{title}{High-speed tunable microwave-rate soliton microcomb}.
\newblock \emph{\bibinfo{journal}{Nature Communications}} \textbf{\bibinfo{volume}{14}}, \bibinfo{pages}{3467} (\bibinfo{year}{2023}).

\bibitem{wang2018integrated}
\bibinfo{author}{Wang, C.} \emph{et~al.}
\newblock \bibinfo{title}{Integrated lithium niobate electro-optic modulators operating at cmos-compatible voltages}.
\newblock \emph{\bibinfo{journal}{Nature}} \textbf{\bibinfo{volume}{562}}, \bibinfo{pages}{101--104} (\bibinfo{year}{2018}).

\bibitem{xu2022dual}
\bibinfo{author}{Xu, M.} \emph{et~al.}
\newblock \bibinfo{title}{Dual-polarization thin-film lithium niobate in-phase quadrature modulators for terabit-per-second transmission}.
\newblock \emph{\bibinfo{journal}{Optica}} \textbf{\bibinfo{volume}{9}}, \bibinfo{pages}{61--62} (\bibinfo{year}{2022}).

\bibitem{wang2018ultrahigh}
\bibinfo{author}{Wang, C.} \emph{et~al.}
\newblock \bibinfo{title}{Ultrahigh-efficiency wavelength conversion in nanophotonic periodically poled lithium niobate waveguides}.
\newblock \emph{\bibinfo{journal}{Optica}} \textbf{\bibinfo{volume}{5}}, \bibinfo{pages}{1438--1441} (\bibinfo{year}{2018}).

\bibitem{lu2019periodically}
\bibinfo{author}{Lu, J.} \emph{et~al.}
\newblock \bibinfo{title}{Periodically poled thin-film lithium niobate microring resonators with a second-harmonic generation efficiency of 250,000\%/w}.
\newblock \emph{\bibinfo{journal}{Optica}} \textbf{\bibinfo{volume}{6}}, \bibinfo{pages}{1455--1460} (\bibinfo{year}{2019}).

\bibitem{mckenna2022ultra}
\bibinfo{author}{McKenna, T.~P.} \emph{et~al.}
\newblock \bibinfo{title}{Ultra-low-power second-order nonlinear optics on a chip}.
\newblock \emph{\bibinfo{journal}{Nature Communications}} \textbf{\bibinfo{volume}{13}}, \bibinfo{pages}{4532} (\bibinfo{year}{2022}).

\bibitem{zhu2021integrated}
\bibinfo{author}{Zhu, D.} \emph{et~al.}
\newblock \bibinfo{title}{Integrated photonics on thin-film lithium niobate}.
\newblock \emph{\bibinfo{journal}{Advances in Optics and Photonics}} \textbf{\bibinfo{volume}{13}}, \bibinfo{pages}{242--352} (\bibinfo{year}{2021}).

\bibitem{zhang2017monolithic}
\bibinfo{author}{Zhang, M.}, \bibinfo{author}{Wang, C.}, \bibinfo{author}{Cheng, R.}, \bibinfo{author}{Shams-Ansari, A.} \& \bibinfo{author}{Lon{\v{c}}ar, M.}
\newblock \bibinfo{title}{Monolithic ultra-high-q lithium niobate microring resonator}.
\newblock \emph{\bibinfo{journal}{Optica}} \textbf{\bibinfo{volume}{4}}, \bibinfo{pages}{1536--1537} (\bibinfo{year}{2017}).

\bibitem{he2019self}
\bibinfo{author}{He, Y.} \emph{et~al.}
\newblock \bibinfo{title}{Self-starting bi-chromatic {L}i{N}b{O}$_{3}$ soliton microcomb}.
\newblock \emph{\bibinfo{journal}{Optica}} \textbf{\bibinfo{volume}{6}}, \bibinfo{pages}{1138--1144} (\bibinfo{year}{2019}).

\bibitem{wan2023photorefraction}
\bibinfo{author}{Wan, S.} \emph{et~al.}
\newblock \bibinfo{title}{Photorefraction-assisted self-emergence of dissipative kerr solitons}.
\newblock \emph{\bibinfo{journal}{arXiv preprint arXiv:2305.02590}}  (\bibinfo{year}{2023}).

\bibitem{carmon2004dynamical}
\bibinfo{author}{Carmon, T.}, \bibinfo{author}{Yang, L.} \& \bibinfo{author}{Vahala, K.~J.}
\newblock \bibinfo{title}{Dynamical thermal behavior and thermal self-stability of microcavities}.
\newblock \emph{\bibinfo{journal}{Optics express}} \textbf{\bibinfo{volume}{12}}, \bibinfo{pages}{4742--4750} (\bibinfo{year}{2004}).

\bibitem{brasch2016photonic}
\bibinfo{author}{Brasch, V.} \emph{et~al.}
\newblock \bibinfo{title}{Photonic chip--based optical frequency comb using soliton cherenkov radiation}.
\newblock \emph{\bibinfo{journal}{Science}} \textbf{\bibinfo{volume}{351}}, \bibinfo{pages}{357--360} (\bibinfo{year}{2016}).

\bibitem{papp2013parametric}
\bibinfo{author}{Papp, S.~B.}, \bibinfo{author}{Del’Haye, P.} \& \bibinfo{author}{Diddams, S.~A.}
\newblock \bibinfo{title}{Parametric seeding of a microresonator optical frequency comb}.
\newblock \emph{\bibinfo{journal}{Optics Express}} \textbf{\bibinfo{volume}{21}}, \bibinfo{pages}{17615--17624} (\bibinfo{year}{2013}).

\bibitem{liu2022ultra}
\bibinfo{author}{Liu, X.} \emph{et~al.}
\newblock \bibinfo{title}{Ultra-broadband and low-loss edge coupler for highly efficient second harmonic generation in thin-film lithium niobate}.
\newblock \emph{\bibinfo{journal}{Advanced Photonics Nexus}} \textbf{\bibinfo{volume}{1}}, \bibinfo{pages}{016001--016001} (\bibinfo{year}{2022}).

\bibitem{zhou2023monolithically}
\bibinfo{author}{Zhou, Y.} \emph{et~al.}
\newblock \bibinfo{title}{Monolithically integrated active passive waveguide array fabricated on thin film lithium niobate using a single continuous photolithography process}.
\newblock \emph{\bibinfo{journal}{Laser \& Photonics Reviews}} \textbf{\bibinfo{volume}{17}}, \bibinfo{pages}{2200686} (\bibinfo{year}{2023}).

\bibitem{zhu2022spectral}
\bibinfo{author}{Zhu, D.} \emph{et~al.}
\newblock \bibinfo{title}{Spectral control of nonclassical light pulses using an integrated thin-film lithium niobate modulator}.
\newblock \emph{\bibinfo{journal}{Light: Science \& Applications}} \textbf{\bibinfo{volume}{11}}, \bibinfo{pages}{327} (\bibinfo{year}{2022}).

\bibitem{jin2021hertz}
\bibinfo{author}{Jin, W.} \emph{et~al.}
\newblock \bibinfo{title}{Hertz-linewidth semiconductor lasers using cmos-ready ultra-high-q microresonators}.
\newblock \emph{\bibinfo{journal}{Nature Photonics}} \textbf{\bibinfo{volume}{15}}, \bibinfo{pages}{346--353} (\bibinfo{year}{2021}).

\bibitem{de2021iii}
\bibinfo{author}{de~Beeck, C.~O.} \emph{et~al.}
\newblock \bibinfo{title}{{III}/{V}-on-lithium niobate amplifiers and lasers}.
\newblock \emph{\bibinfo{journal}{Optica}} \textbf{\bibinfo{volume}{8}}, \bibinfo{pages}{1288--1289} (\bibinfo{year}{2021}).

\bibitem{shams2022electrically}
\bibinfo{author}{Shams-Ansari, A.} \emph{et~al.}
\newblock \bibinfo{title}{Electrically pumped laser transmitter integrated on thin-film lithium niobate}.
\newblock \emph{\bibinfo{journal}{Optica}} \textbf{\bibinfo{volume}{9}}, \bibinfo{pages}{408--411} (\bibinfo{year}{2022}).

\bibitem{li2022integrated}
\bibinfo{author}{Li, M.} \emph{et~al.}
\newblock \bibinfo{title}{Integrated pockels laser}.
\newblock \emph{\bibinfo{journal}{Nature communications}} \textbf{\bibinfo{volume}{13}}, \bibinfo{pages}{5344} (\bibinfo{year}{2022}).

\bibitem{guo2023ultrafast}
\bibinfo{author}{Guo, Q.} \emph{et~al.}
\newblock \bibinfo{title}{Ultrafast mode-locked laser in nanophotonic lithium niobate}.
\newblock \emph{\bibinfo{journal}{Science}} \textbf{\bibinfo{volume}{382}}, \bibinfo{pages}{708--713} (\bibinfo{year}{2023}).

\bibitem{snigirev2023ultrafast}
\bibinfo{author}{Snigirev, V.} \emph{et~al.}
\newblock \bibinfo{title}{Ultrafast tunable lasers using lithium niobate integrated photonics}.
\newblock \emph{\bibinfo{journal}{Nature}} \textbf{\bibinfo{volume}{615}}, \bibinfo{pages}{411--417} (\bibinfo{year}{2023}).

\bibitem{del2009frequency}
\bibinfo{author}{Del'Haye, P.}, \bibinfo{author}{Arcizet, O.}, \bibinfo{author}{Gorodetsky, M.~L.}, \bibinfo{author}{Holzwarth, R.} \& \bibinfo{author}{Kippenberg, T.~J.}
\newblock \bibinfo{title}{Frequency comb assisted diode laser spectroscopy for measurement of microcavity dispersion}.
\newblock \emph{\bibinfo{journal}{Nature photonics}} \textbf{\bibinfo{volume}{3}}, \bibinfo{pages}{529--533} (\bibinfo{year}{2009}).

\bibitem{brasch2017self}
\bibinfo{author}{Brasch, V.}, \bibinfo{author}{Lucas, E.}, \bibinfo{author}{Jost, J.~D.}, \bibinfo{author}{Geiselmann, M.} \& \bibinfo{author}{Kippenberg, T.~J.}
\newblock \bibinfo{title}{Self-referenced photonic chip soliton kerr frequency comb}.
\newblock \emph{\bibinfo{journal}{Light: Science \& Applications}} \textbf{\bibinfo{volume}{6}}, \bibinfo{pages}{e16202--e16202} (\bibinfo{year}{2017}).

\bibitem{Bhaskaran:2021:Nature}
\bibinfo{author}{Feldmann, J.} \emph{et~al.}
\newblock \bibinfo{title}{Parallel convolutional processing using an integrated photonic tensor core}.
\newblock \emph{\bibinfo{journal}{Nature}} \textbf{\bibinfo{volume}{589}}, \bibinfo{pages}{52--58} (\bibinfo{year}{2021}).

\bibitem{Moss:2021:Nature}
\bibinfo{author}{Xu, X.} \emph{et~al.}
\newblock \bibinfo{title}{11 tops photonic convolutional accelerator for optical neural networks}.
\newblock \emph{\bibinfo{journal}{Nature}} \textbf{\bibinfo{volume}{589}}, \bibinfo{pages}{44--51} (\bibinfo{year}{2021}).

\bibitem{he2019low}
\bibinfo{author}{He, L.} \emph{et~al.}
\newblock \bibinfo{title}{Low-loss fiber-to-chip interface for lithium niobate photonic integrated circuits}.
\newblock \emph{\bibinfo{journal}{Optics letters}} \textbf{\bibinfo{volume}{44}}, \bibinfo{pages}{2314--2317} (\bibinfo{year}{2019}).

\bibitem{yuan2023soliton}
\bibinfo{author}{Yuan, Z.} \emph{et~al.}
\newblock \bibinfo{title}{Soliton pulse pairs at multiple colors in normal dispersion microresonators}.
\newblock \emph{\bibinfo{journal}{arXiv preprint arXiv:2301.10976}}  (\bibinfo{year}{2023}).

\bibitem{xue2015mode}
\bibinfo{author}{Xue, X.} \emph{et~al.}
\newblock \bibinfo{title}{Mode-locked dark pulse kerr combs in normal-dispersion microresonators}.
\newblock \emph{\bibinfo{journal}{Nature Photonics}} \textbf{\bibinfo{volume}{9}}, \bibinfo{pages}{594--600} (\bibinfo{year}{2015}).

\bibitem{helgason2023surpassing}
\bibinfo{author}{Helgason, {\'O}.~B.} \emph{et~al.}
\newblock \bibinfo{title}{Surpassing the nonlinear conversion efficiency of soliton microcombs}.
\newblock \emph{\bibinfo{journal}{Nature Photonics}} \textbf{\bibinfo{volume}{17}}, \bibinfo{pages}{992--999} (\bibinfo{year}{2023}).

\bibitem{jung2021tantala}
\bibinfo{author}{Jung, H.} \emph{et~al.}
\newblock \bibinfo{title}{Tantala kerr nonlinear integrated photonics}.
\newblock \emph{\bibinfo{journal}{Optica}} \textbf{\bibinfo{volume}{8}}, \bibinfo{pages}{811--817} (\bibinfo{year}{2021}).

\bibitem{yi2015soliton}
\bibinfo{author}{Yi, X.}, \bibinfo{author}{Yang, Q.-F.}, \bibinfo{author}{Yang, K.~Y.}, \bibinfo{author}{Suh, M.-G.} \& \bibinfo{author}{Vahala, K.}
\newblock \bibinfo{title}{Soliton frequency comb at microwave rates in a high-q silica microresonator}.
\newblock \emph{\bibinfo{journal}{Optica}} \textbf{\bibinfo{volume}{2}}, \bibinfo{pages}{1078--1085} (\bibinfo{year}{2015}).

\bibitem{wu2023algaas}
\bibinfo{author}{Wu, L.} \emph{et~al.}
\newblock \bibinfo{title}{Algaas soliton microcombs at room temperature}.
\newblock \emph{\bibinfo{journal}{Optics Letters}} \textbf{\bibinfo{volume}{48}}, \bibinfo{pages}{3853--3856} (\bibinfo{year}{2023}).

\bibitem{gong2020near}
\bibinfo{author}{Gong, Z.}, \bibinfo{author}{Liu, X.}, \bibinfo{author}{Xu, Y.} \& \bibinfo{author}{Tang, H.~X.}
\newblock \bibinfo{title}{Near-octave lithium niobate soliton microcomb}.
\newblock \emph{\bibinfo{journal}{Optica}} \textbf{\bibinfo{volume}{7}}, \bibinfo{pages}{1275--1278} (\bibinfo{year}{2020}).

\bibitem{guidry2022quantum}
\bibinfo{author}{Guidry, M.~A.}, \bibinfo{author}{Lukin, D.~M.}, \bibinfo{author}{Yang, K.~Y.}, \bibinfo{author}{Trivedi, R.} \& \bibinfo{author}{Vu{\v{c}}kovi{\'c}, J.}
\newblock \bibinfo{title}{Quantum optics of soliton microcombs}.
\newblock \emph{\bibinfo{journal}{Nature Photonics}} \textbf{\bibinfo{volume}{16}}, \bibinfo{pages}{52--58} (\bibinfo{year}{2022}).

\bibitem{hausmann2014diamond}
\bibinfo{author}{Hausmann, B.}, \bibinfo{author}{Bulu, I.}, \bibinfo{author}{Venkataraman, V.}, \bibinfo{author}{Deotare, P.} \& \bibinfo{author}{Lon{\v{c}}ar, M.}
\newblock \bibinfo{title}{Diamond nonlinear photonics}.
\newblock \emph{\bibinfo{journal}{Nature Photonics}} \textbf{\bibinfo{volume}{8}}, \bibinfo{pages}{369--374} (\bibinfo{year}{2014}).

\bibitem{wilson2020integrated}
\bibinfo{author}{Wilson, D.~J.} \emph{et~al.}
\newblock \bibinfo{title}{Integrated gallium phosphide nonlinear photonics}.
\newblock \emph{\bibinfo{journal}{Nature Photonics}} \textbf{\bibinfo{volume}{14}}, \bibinfo{pages}{57--62} (\bibinfo{year}{2020}).

\bibitem{moille2023fourier}
\bibinfo{author}{Moille, G.}, \bibinfo{author}{Lu, X.}, \bibinfo{author}{Stone, J.}, \bibinfo{author}{Westly, D.} \& \bibinfo{author}{Srinivasan, K.}
\newblock \bibinfo{title}{Fourier synthesis dispersion engineering of photonic crystal microrings for broadband frequency combs}.
\newblock \emph{\bibinfo{journal}{Communications Physics}} \textbf{\bibinfo{volume}{6}}, \bibinfo{pages}{144} (\bibinfo{year}{2023}).

\bibitem{yu2021spontaneous}
\bibinfo{author}{Yu, S.-P.} \emph{et~al.}
\newblock \bibinfo{title}{Spontaneous pulse formation in edgeless photonic crystal resonators}.
\newblock \emph{\bibinfo{journal}{Nature Photonics}} \textbf{\bibinfo{volume}{15}}, \bibinfo{pages}{461--467} (\bibinfo{year}{2021}).

\bibitem{lucas2023tailoring}
\bibinfo{author}{Lucas, E.}, \bibinfo{author}{Yu, S.-P.}, \bibinfo{author}{Briles, T.~C.}, \bibinfo{author}{Carlson, D.~R.} \& \bibinfo{author}{Papp, S.~B.}
\newblock \bibinfo{title}{Tailoring microcombs with inverse-designed, meta-dispersion microresonators}.
\newblock \emph{\bibinfo{journal}{Nature Photonics}} \textbf{\bibinfo{volume}{17}}, \bibinfo{pages}{943--950} (\bibinfo{year}{2023}).

\bibitem{yang2017stokes}
\bibinfo{author}{Yang, Q.-F.}, \bibinfo{author}{Yi, X.}, \bibinfo{author}{Yang, K.~Y.} \& \bibinfo{author}{Vahala, K.}
\newblock \bibinfo{title}{Stokes solitons in optical microcavities}.
\newblock \emph{\bibinfo{journal}{Nature Physics}} \textbf{\bibinfo{volume}{13}}, \bibinfo{pages}{53--57} (\bibinfo{year}{2017}).

\bibitem{bao2019laser}
\bibinfo{author}{Bao, H.} \emph{et~al.}
\newblock \bibinfo{title}{Laser cavity-soliton microcombs}.
\newblock \emph{\bibinfo{journal}{Nature Photonics}} \textbf{\bibinfo{volume}{13}}, \bibinfo{pages}{384--389} (\bibinfo{year}{2019}).

\bibitem{bruch2021pockels}
\bibinfo{author}{Bruch, A.~W.} \emph{et~al.}
\newblock \bibinfo{title}{Pockels soliton microcomb}.
\newblock \emph{\bibinfo{journal}{Nature Photonics}} \textbf{\bibinfo{volume}{15}}, \bibinfo{pages}{21--27} (\bibinfo{year}{2021}).

\bibitem{cole2017soliton}
\bibinfo{author}{Cole, D.~C.}, \bibinfo{author}{Lamb, E.~S.}, \bibinfo{author}{Del’Haye, P.}, \bibinfo{author}{Diddams, S.~A.} \& \bibinfo{author}{Papp, S.~B.}
\newblock \bibinfo{title}{Soliton crystals in kerr resonators}.
\newblock \emph{\bibinfo{journal}{Nature Photonics}} \textbf{\bibinfo{volume}{11}}, \bibinfo{pages}{671--676} (\bibinfo{year}{2017}).

\bibitem{stern2018battery}
\bibinfo{author}{Stern, B.}, \bibinfo{author}{Ji, X.}, \bibinfo{author}{Okawachi, Y.}, \bibinfo{author}{Gaeta, A.~L.} \& \bibinfo{author}{Lipson, M.}
\newblock \bibinfo{title}{Battery-operated integrated frequency comb generator}.
\newblock \emph{\bibinfo{journal}{Nature}} \textbf{\bibinfo{volume}{562}}, \bibinfo{pages}{401--405} (\bibinfo{year}{2018}).

\bibitem{shen2020integrated}
\bibinfo{author}{Shen, B.} \emph{et~al.}
\newblock \bibinfo{title}{Integrated turnkey soliton microcombs}.
\newblock \emph{\bibinfo{journal}{Nature}} \textbf{\bibinfo{volume}{582}}, \bibinfo{pages}{365--369} (\bibinfo{year}{2020}).

\bibitem{kim2019turn}
\bibinfo{author}{Kim, B.~Y.} \emph{et~al.}
\newblock \bibinfo{title}{Turn-key, high-efficiency kerr comb source}.
\newblock \emph{\bibinfo{journal}{Optics letters}} \textbf{\bibinfo{volume}{44}}, \bibinfo{pages}{4475--4478} (\bibinfo{year}{2019}).

\bibitem{raja2019electrically}
\bibinfo{author}{Raja, A.~S.} \emph{et~al.}
\newblock \bibinfo{title}{Electrically pumped photonic integrated soliton microcomb}.
\newblock \emph{\bibinfo{journal}{Nature communications}} \textbf{\bibinfo{volume}{10}}, \bibinfo{pages}{680} (\bibinfo{year}{2019}).

\end{thebibliography}
\newpage

\section{Methods}

\noindent\textbf{Device fabrication and parameters}   \\
Dissipative Kerr soliton (DKS) devices are fabricated on a Z-cut thin-film lithium niobate (Z-TFLN) on insulator wafer supplied by NanoLN. The initial film thickness is 600 nm of Z-TFLN on top of 2 $\mu$m thermal oxide and 0.525 mm silicon. Nanophotonic waveguides are patterned on hydrogen silsesquioxane (HSQ) resist using electron-beam lithography (EBL). The patterned resist undergoes multiple iterations of Ar$^+$-based reactive ion etching and wet etching until an etch depth of about 345 nm is reached, leaving about 255 nm of slab. This iterative etching process results in a waveguide sidewall angle of about 72 degrees, owing to the prevention of excess redeposition buildup during the dry-etching process. Finally, the HSQ resist is stripped using dilute hydrogen-fluoride (HF) and the devices are annealed in a high temperature, oxygen-rich environment. Routing waveguides for coupling light on and off the chip are exposed through manual cleaving, resulting in coupling losses of about 6 dB per facet.

Electro-optic (EO) phase modulator   devices are fabricated on a X-cut thin-film lithium niobate (X-TFLN) on insulator wafer supplied by NanoLN. The initial film thickness is 600 nm of X-TFLN on top of 2 $\mu$m thermal oxide and 0.525 mm silicon. The fabrication process is identical to that for DKS devices up to etching of the waveguides (320 nm etch depth and 280 nm slab, instead). Plasma-enhanced chemical vapor deposition (PECVD) is used to deposit a silica top cladding layer. Bilayer taper mode converters are fabricated by first opening photolithography-defined rectangular windows in the top cladding using HF wet etching. Subsequently, the mode converter tips are defined using aligned EBL and another round of Ar$^+$-based reactive ion etching. Finally, the devices are cleaned and cladded with PECVD silica once again. Two centimeter long gold microwave electrodes are defined near the waveguides using aligned EBL and photolithography and metalized using electron-beam deposition followed by liftoff process. Facets at the mode converter tips are exposed through reactive ion etching through the X-TFLN, thermal oxide, and silicon, resulting in a total EO phase modulator insertion loss of about 8 dB. The 6 GHz half-wave-voltage ($V_\pi$) is about 4 V when optically pumped in the C-band.

\vspace{1ex}\noindent\textbf{Microresonator dispersion engineering} \\
The microresonator dispersion is defined by the integrated dispersion $D_{\text{int}}(\mu)=\omega_{\mu}-\omega_{0}-D_{1}\mu=\sum_{n\geq2}\frac{D_n}{n!}\mu^{n}$, where $\mu$ is the azimuthal mode index\cite{brasch2016photonic}. The dispersion parameters $D_n$ are directly obtained from eigenmode simulations of the fundamental transverse electric (TE) mode effective refractive index ($n_{\text{eff}}$), directly due to engineered microresonator waveguide cross-sections. We carry out such simulations using a commercial eigenmode solver (Lumerical MODE) including waveguide bending. We present the simulation of $D_{\text{int}}$ for all DKS states in Fig. 2\textbf{e}-\textbf{g}, alongside representative fundamental Transverse Electric (TE) modal profiles for various waveguide cross-sectional geometries in Extended Fig. 1. In these figures, the waveguide width is varied across different microresonator radii, with the waveguide height (etch depth) consistently maintained at approximately 345 nm. This effectively illustrates how the positions of dispersive waves are influenced by the waveguide width parameter. In Fig. 2\textbf{c}, the experimental $D_{\text{int}}$ in a narrow band around the pump wavelength (1520-1630 nm) is measured by fitting the measured resonance positions of the microresonator, where their frequencies are calibrated by a fiber-based Mach-Zehnder interferometer with a fringe period of 191.3 MHz. The good agreement between simulated and experimental dispersive wave positions and $D_{\text{int}}$ suggests predictable dispersion engineering for broadband DKS states in the TFLN platform. 

\vspace{1ex}\noindent\textbf{Dissipative Kerr soliton device characterization} \\
The experimental setup for DKS device characterization and soliton generation is schematically illustrated in Extended Fig. 2. A continuous-wave (CW) pump laser is amplified by an erbium-doped fiber amplifier (EDFA). In Fig. 2\textbf{c}, a tunable bandpass filter was used to filter out the amplified spontaneous emission (ASE) associated noise with the pump power amplification. Further, a fiber-Bragg-grating (FBG) notch filter filters out the strong pump frequency component in the DKS spectrum. In Fig. 2\textbf{d}-\textbf{g}, the ASE was not removed from the spectrum and corresponds to the irregular shape beneath the DKS envelope. The pump was also not removed due to over ten-meters of compressed fiber in the notch filter, introducing excess fiber-induced losses. In both cases, polarization controllers and lensed fibers are used to couple light into the DKS chip. Lensed fibers are used to couple light off the DKS chip. One percent of the total output is used to monitor the fiber-to-chip in and out coupling. Ninety-nine percent is then split such that ten percent is sent to a 125 MHz photoreceiver to monitor the total comb power as a voltage readout on an oscilloscope and ninety percent is sent to two optical spectrum analyzers (OSAs, covering 600-1700 nm and 1200-2400 nm) to simultaneously monitor the comb spectrum. The comb power measurement during laser scanning back and forth through a microresonator resonance (Fig. 2\textbf{a}) is recorded by the oscilloscope voltage. The stability measurement (Fig. 2\textbf{b}), as the DKS state is maintained, consists of oscilloscope voltage values collected every 20 seconds over a total of 15 hours.

\vspace{1ex}\noindent\textbf{Electro-optic phase modulator performance characterization} \\
The experimental setup for electro-optic (EO) phase modulator performance characterization is schematically illustrated in Extended Fig. 2. A CW pump laser passes through a polarization controller and is coupled on and off the EO phase modulator chip by lensed fibers. One percent of the total output is used to monitor the fiber-to-chip in and out coupling and polarization. Ninety-nine percent is sent to an OSA (covering 600-1700 nm) to monitor the EO comb spectrum, such as those in Fig. 3\textbf{a}, where five CW optical frequencies are sequentially coupled. The electrical power to generate these spectra is estimated to be about 4.47 W (36.5 dBm). This electrical power was chosen solely to demonstrate the power-handling capabilities of our modulator chip and a lower power level was used for hybrid Kerr-EO comb generation. Since the EO phase modulator is required to operate at all frequencies covering the source DKS span, our integrated phase modulator offers the uniquely existing on-chip solution while also being highly compact and power stable. We characterized its 6 GHz half-wave voltage ($V_\pi$) and the electro-optic bandwidth at fifteen optical wavelengths in the telecommunications L, C, S, and E bands, only limited by available lasers, using the methodology described below. At each wavelength, 137 modulation frequencies (6 GHz to 40 GHz in steps of 0.25 GHz) are applied at power levels of 5 and 10 dBm output directly from a calibrated microwave source. For each modulation frequency and source output power combination, the EO comb spectrum is collected and the power in each comb line extracted. The power of the $n^{\text{th}}$ EO sideband theoretically corresponds to the nth order Bessel function $J_n(\beta_m)$ evaluated at a modulation depth given by $\beta_m=\pi V/V_\pi$. Since the power delivered onto the modulator gives $V$, we fit the EO sideband powers and delivered microwave power (source output power corrected for microwave circuit losses) to obtain the $V_\pi$ at some modulation frequency and source output power combination. A single-valued $V_\pi$ for a given modulation frequency was obtained by averaging the two $V_\pi$s extracted when the source output power was varied between 5 and 10 dBm. Averaging over more power levels was not necessary as they always yielded near identical $V_\pi$, which further justifies averaging as only a means of reducing statistical variation in $V_\pi$ measurements at the same modulation frequency. Specifically, we repeated this $V_\pi$ vs. modulation frequency measurement at fifteen optical wavelengths of 1355 nm and 1370-1630 nm in intervals of 20 nm, and three representative curves were shown in Fig. 3\textbf{b}. The first value of each curve is taken to be the 6 GHz (plotted as data points in Fig. 3\textbf{c}), and the modulation frequency at which the $V_\pi$ rises to $\sqrt{2}$ of its 6 GHz value is taken to be the EO 3 dB bandwidth (plotted as data points in Fig. 3\textbf{d}).

\vspace{1ex}\noindent\textbf{Hybrid Kerr-electro-optic comb generation} \\
The experimental setup for hybrid Kerr-EO comb generation is schematically illustrated in Extended Fig. 2. It is a combination of the separate setups for DKS generation and EO phase modulator characterization, except the two chips are operated simultaneously, with an FBG notch filter to filter out the strong pump frequency component and a polarization controller linking the chips. Two OSAs are used to simultaneously monitor the comb spectrum. The measurement did not necessitate use of the FBG, and the polarization controller may be omitted in conceivable monolithic integration or low-loss coupling schemes between Z-TFLN and X-TFLN, due to the good spatial mode overlap between the fundamental TE modes in both cuts of thin-film material.

To generate the hybrid Kerr-EO combs as in Fig. 4a, the optical power on-chip was estimated to be 125 mW from an EDFA output power of 500 mW, accounting for a 6 dB coupling loss induced by the input facet. Notably, with bilayer taper mode converters such as those fabricated on the EO phase modulator chip, the coupling loss may be reduced to 1.7 dB per facet and the off-chip pump power requirement lowered to 188 mW\cite{he2019low}. Advanced coupler designs may be exploited for 0.54 dB loss per facet and the off-chip pump power requirement lowered to 142 mW\cite{liu2022ultra}. Currently, the loaded quality factor ($Q_L$) for our pump resonance in Fig. 4a is about 1 million. Fabrication improvements in $Q_L$ by a factor of 2 may reduce the on-chip pump power requirement. Packaged distributed feedback (DFB) lasers can be butt-coupled to the DKS chip facet in combination with edge couplers to yield estimated facet losses conservatively around 3 dB. 

The electrical power required to electro-optically divide 410.319 GHz of DKS spacing into microwave-rate (29.308 GHz) separated lines was calibrated to be 2.51 W (34.5 dBm) after accounting for microwave losses in the system, at 29.158 GHz modulation frequency. We specifically selected a microwave power such that the highest power sidebands would appear at the $N/2=7$ sideband. This choice ensures that the comb lines separated by the difference frequency $\Delta f$ would generate a microwave beatnote with high signal-to-noise ratio and phase locking $\Delta f$ to the stable microwave oscillator would yield a high-quality phase lock in the locking demonstration. The EO phase modulator measured in the C-band has a $V_\pi L\sim8.84 V\cdot\text{cm}$ at 6 GHz modulation frequency and a 3 dB EO bandwidth of 31.47 GHz. The electrical power consumption may be further lowered by 5.3 dB using state-of-the-art TFLN EO phase modulators with $V_\pi L\sim4.8 V\cdot\text{cm}$ (extrapolated from quoted $2.4 V\cdot\text{cm}$ for a dual-drive amplitude modulator)\cite{xu2022dual} and 110 GHz 3 dB EO bandwidths. This projects an electrical power on-chip of 0.24 W (23.8 dBm), reaching sub-Watt levels. Dual-drive phase modulators in a loop-back architecture may be realized\cite{zhu2022spectral}, further lowering the projected electrical power on-chip by a factor of two, down to 0.12 W (20.8 dBm).

\vspace{1ex}\noindent\textbf{Soliton spacing detection, locking, and stabilization} \\
When generating the hybrid Kerr-EO comb, a nonzero difference frequency $\Delta f$ is introduced when $f_{DKS}$ is not an integer multiple of $f_{RF}$. In our experiment in Fig. 4d (setup schematically illustrated in Extended Fig. 3), $\Delta f$ was locked at 2.107 GHz after detection by a 12 GHz bandwidth photoreceiver when $f_{RF}$ was set to 29.158 GHz. The lock was maintained by electronic feedback onto the current control of the pump laser, controlling its frequency while residual laser intensity fluctuations are negligible. During this lock, $f_{DKS}$  was fixed at 410.319 GHz. To generate the $\Delta f$ beatnote on the photoreceiver, relevant EO sideband pairs were filtered then amplified by a C-band pre-amplifier before beating on the photoreceiver. The $\Delta f$ was further amplified by a low-noise microwave amplifier (providing about 26 dB of gain) to produce a sufficient beatnote power level for the phase locked loop (PLL). The PLL is based on a phase comparator circuit where the reference frequency is 32 times a signal generator set at 65.865 MHz. This signal generator is externally referenced by the 10 kHz clock signal synthesizing $f_{RF}$. In principle, the broadband, microwave-rate frequency reference provided by our hybrid Kerr-EO comb may be fully stabilized, that is, the frequency of each of the 2,589 comb lines can be precisely known, once both $f_{ceo}$ and $f_{DKS}$ of the frequency comb are stabilized. Here, we demonstrated the possibility for the detection, stabilization, and control of a near THz-rate $f_{DKS}$ using a single integrated EO phase modulator.

\vspace{1ex}\noindent\textbf{Comparison with other integrated frequency comb technologies} \\
Comparing our work with state-of-the-art integrated frequency comb technologies surveying various material platforms, such as octave-spanning DKS combs and microresonator-based EO-combs, we find that the hybrid Kerr-EO comb of this work produces the largest span (75.9 THz) while simultaneously capable of directly interfacing with conventional fast-electronics, which we define as a sub-50 GHz spacing between adjacent comb lines. These features are augmented by a record number of 2,589 lines produced using a comb-generator consisting of purely integrated components. This comparison is graphically illustrated in Extended Fig. 4. All data points are directly quoted from references or estimated from figures when exact numbers are not reported\cite{zhang2019broadband, hu2022high, he2023high, moille2023kerr, li2017stably, pfeiffer2017octave, yuan2023soliton, xue2015mode, helgason2023surpassing, jung2021tantala, liu2021aluminum, weng2021directly, yi2015soliton, wu2023algaas, gong2020near, guidry2022quantum, hausmann2014diamond, wilson2020integrated}. We note that significant achievements were made in integrated frequency combs beyond DKS or EO combs in the most conventional sense, such as ones utilizing advanced dispersion engineering\cite{moille2023fourier, yu2021spontaneous,lucas2023tailoring} and additional nonlinear optical effects\cite{yang2017stokes,bao2019laser,bruch2021pockels}, soliton crystals\cite{cole2017soliton}, and lower operational power and turnkey microcombs\cite{stern2018battery,shen2020integrated,kim2019turn,raja2019electrically}, which cannot be simply viewed by span, spacing, or number of comb lines.\\

\noindent\textbf{Data availability}  The data that support the plots within this paper and other findings of this study are available from the corresponding author upon reasonable request.

\noindent\textbf{Code availability}  The code used to produce the plots within this paper is available from the corresponding author upon reasonable request.

\section*{Acknowledgments}
This work is supported by the Defense Advanced Research Projects Agency (D23AP00251-00, HR001120C0137), Office of Naval Research (N00014-22-C-1041), National Science Foundation (OMA-2137723, OMA-2138068), and National Research Foundation of Korea. The authors thank Rebecca Cheng for providing the phase modulator, Chaoshen Zhang, Hana Warner, Rebecca Cheng, and Neil Sinclair for discussions, and Let\'icia Magalh\~aes and Jeremiah Jacobson for photography assistance. Y.S. acknowledges support from the AWS Generation Q Fund at the Harvard Quantum Initiative. All devices in this work were fabricated at the Harvard Center for Nanoscale Systems.

\clearpage
\renewcommand{\figurename}{Extended Fig.}
\renewcommand{\thefigure}{\arabic{figure}}
\setcounter{figure}{0}

\begin{figure*}[t!]
\centering
\includegraphics[width=\linewidth]{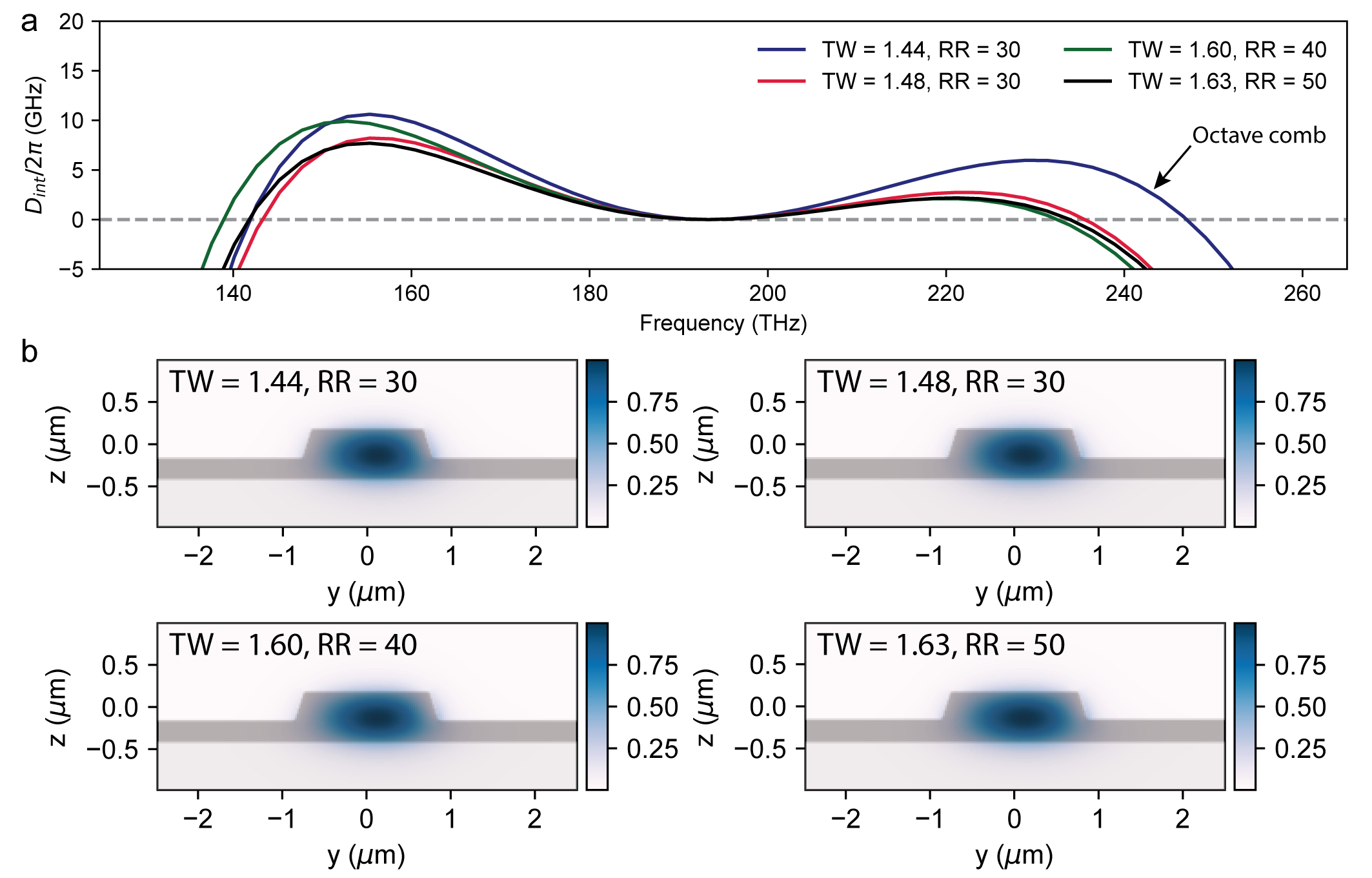}
\caption{\label{fig:EDFig1}\textbf{Dispersion engineering of dissipative Kerr soliton on thin-film lithium niobate.} \textbf{a} Simulated integrated dispersion $D_{int}/2\pi$ of four representative microresonators in Fig. 2d-g of the main text. In these cases, the waveguide parameters are tuned such that $D_{int}/2\pi$ are similar and correspond to sufficiently broadband DKS states. Fig. 2d and 2e bottom (blue); Fig. 2e top (red); Fig. 2f top (green); and Fig. 2g top (black). \textbf{b} Simulated $E_y$ of the fundamental transerve electric mode electric field distribution, illustrating the effects of waveguide bending and modal confinement on its shape and hence the effective index.}
\end{figure*}

\begin{figure*}[t!]
\centering
\includegraphics[width=\linewidth]{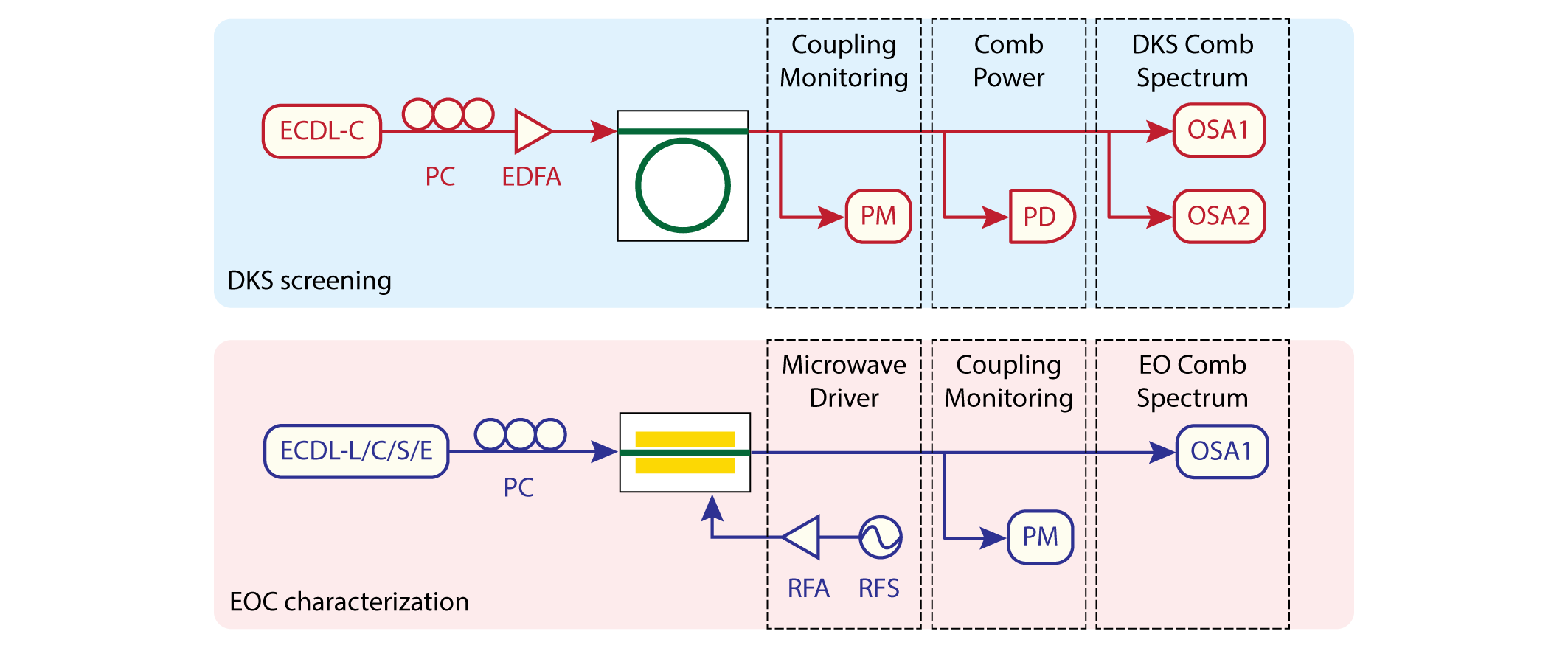}
\caption{\label{fig:EDFig2}\textbf{Dissipative Kerr soliton screening and electro-optic phase modulator characterization setups.} Blue (red) panel is the setup schematic for DKS screening (EO modulator characterization). ECDL-L/C/S/E: external cavity diode laser in the L/C/S/E-band. PC: polarization control paddles. EDFA: erbium-doped fiber amplifier. PM: power monitor unit. PD: photoreceiver. OSA1/2: 1200-2400 nm and 600-1700 nm spectral coverage optical spectrum analyzers. RFS: microwave source. RFA: microwave amplifier.}
\end{figure*}

\begin{figure*}[t!]
\centering
\includegraphics[width=\linewidth]{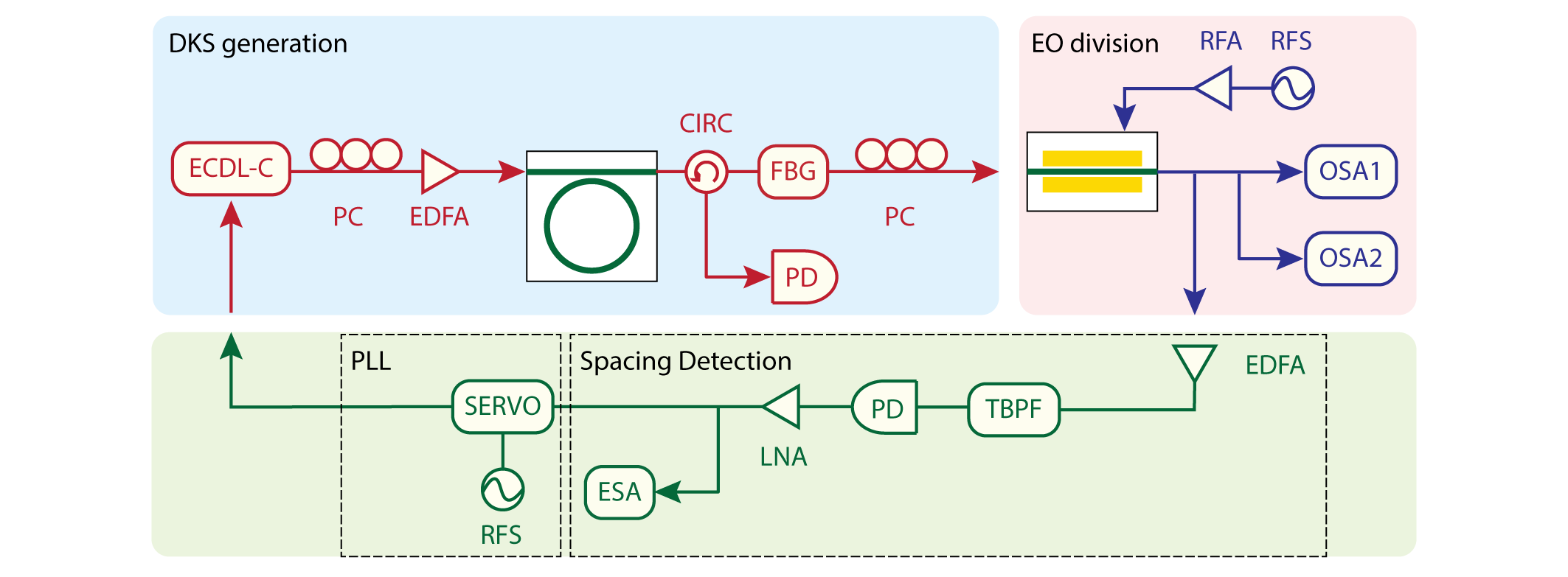}
\caption{\label{fig:EDFig3}\textbf{Hybrid Kerr-electro-optic comb generation and stabilization setup.} Blue panel generates the DKS comb. The DKS comb is fed into the red panel for EO division, which divides the near THz-rate spacing down to microwave rates. The hybrid Kerr-EO comb is fed into the green panel, which detects the difference frequency $\Delta f$ between the $\pm$7 order sidebands of adjacent DKS comb lines, phase locks $f_o$ with a stable microwave oscillator, producing an error signal fed back to the pump current for pump frequency adjustment. ECDL-C: external cavity diode laser in the C-band. PC: polarization control paddles. EDFA: erbium-doped fiber amplifier. CIRC: circulator. FBG: fiber Bragg grating. PD: photoreceiver. OSA1/2: 1200-2400 nm and 600-1700 nm spectral coverage optical spectrum analyzers. RFS: microwave source. RFA: microwave amplifier. TBPF: tunable bandpass filter. LNA: low-noise microwave amplifier. ESA: electrical spectrum analyzer. SERVO: electronic servo control. PLL: phase locked loop.}
\end{figure*}

\begin{figure*}[t!]
\centering
\includegraphics[width=\linewidth]{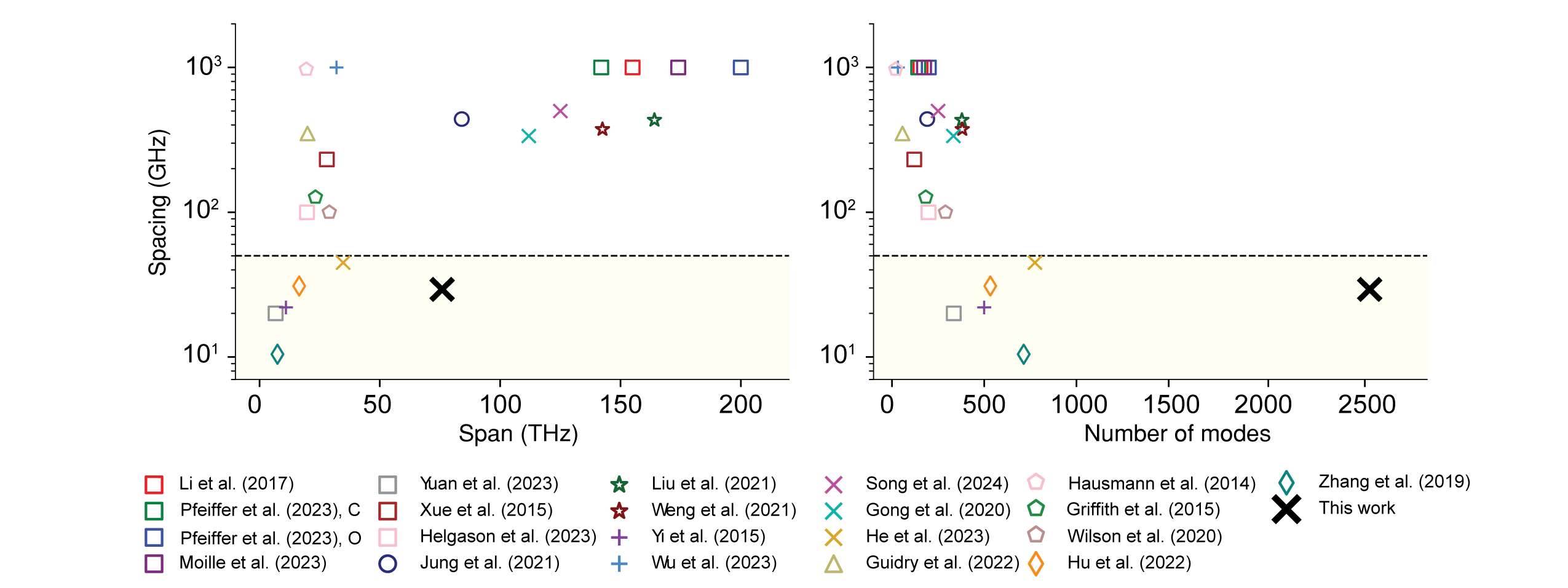}
\caption{\label{fig:EDFig4}\textbf{Comparison with other integrated optical frequency comb technologies.} Our hybrid Kerr-electro-optic frequency comb (black cross) placed in the landscape other fully integrated frequency comb generators, considering spacing, comb span, and number of modes. The black dashed line marks a spacing of 50 GHz and shaded yellow region represents region interfaced with conventional fast electronics.}
\end{figure*}

\end{document}